\documentclass[preprint,12pt]{elsarticle}

\usepackage{amssymb}

\usepackage{algorithm}
\usepackage{algpseudocode}  
\usepackage[utf8]{inputenc}

\newcounter{bla}

\journal{Computer Physics Communications}

\usepackage{siunitx}
\usepackage{graphicx} 
\usepackage[utf8]{inputenc}
\usepackage[T1]{fontenc}
\usepackage{amsmath}
\usepackage{amsfonts}
\usepackage{amssymb}
\usepackage{array}
\usepackage[version=3]{mhchem} 
\usepackage{braket}
\usepackage{bm}
\usepackage{xr}
\usepackage{xspace}
\usepackage{multirow}
\usepackage{cleveref}
\usepackage{titlesec}
\usepackage{xspace} 
\usepackage{caption}
\usepackage{subcaption}
\usepackage{tikz}
\usetikzlibrary{arrows, arrows.meta}
\usepackage{comment}
\usepackage{newtxtext}
\usepackage{mathtools}
\usepackage{fancyvrb}
\usepackage{framed}
\usepackage{xcolor}
\usepackage{adjustbox}
\usepackage{comment}

\DeclareUnicodeCharacter{03BC}{\ensuremath{\mu}}
\DeclareUnicodeCharacter{03C9}{\ensuremath{\omega}}
\DeclareUnicodeCharacter{0212}{\ensuremath{\AA}}

\definecolor{shadecolor}{gray}{0.95}

\newcommand*{\wfq}{$\omega$FQ\xspace}
\newcommand*{\wfqfmu}{$\omega$FQF$\mu$\xspace}

\newcommand*{\imm}{\mathrm{i}}

\newcommand*{\tqq}{\mathbf{T}^\mathrm{qq}}
\newcommand*{\tqmu}{\mathbf{T}^\mathrm{q\mu}}
\newcommand*{\tmuq}{\mathbf{T}^\mathrm{\mu q}}
\newcommand*{\tmumu}{\mathbf{T}^\mathrm{\mu\mu}}

\begin{document}

\begin{frontmatter}



\title{plasmonX: an Open-Source Code for Nanoplasmonics}

\author[a]{Tommaso Giovannini\corref{author1}}
\author[b]{Pablo Grobas Illobre}
\author[b]{Piero Lafiosca}
\author[b]{Luca Nicoli}
\author[b]{Luca Bonatti}
\author[c,d]{Stefano Corni}
\author[b]{Chiara Cappelli\corref{author2}}

\cortext[author1] {Corresponding author.\\\textit{E-mail address:} tommaso.giovannini@uniroma2.it}
\cortext[author2] {Corresponding author.\\\textit{E-mail address:} chiara.cappelli@sns.it}
\address[a]{Department of Physics and INFN, University of Rome Tor Vergata, Via della Ricerca Scientifica, 1, 00133, Rome, Italy}
\address[b]{Scuola Normale Superiore, Piazza dei Cavalieri, 7, 56126 Pisa, Italy}
\address[c]{Department of Chemical Sciences, University of Padova, via Marzolo 1, Padova, Italy}
\address[d]{CNR Institute of Nanoscience, via Campi 213/A, Modena, Italy}

\begin{abstract}
We present the first public release of \texttt{plasmonX}, a novel open-source code for simulating the plasmonic response of complex nanostructures. The code supports both fully atomistic and implicit descriptions of nanomaterials. In particular, it employs the frequency-dependent fluctuating charges ($\omega$FQ) and dipoles ($\omega$FQF$\mu$) models to describe the response properties of atomistic structures, including simple and $d$-metals, graphene-based structures, and multi-metal nanostructures. For implicit representations, the Boundary Element Method is implemented in both the dielectric polarizable continuum model (DPCM) and integral equation formalism (IEF-PCM) variants. The distribution also includes a post-processing module that enables analysis of electric field-induced properties such as charge density and electric field patterns. 

\noindent \textbf{PROGRAM SUMMARY}

\begin{small}
\noindent
{\em Program Title:} plasmonX \\
{\em CPC Library link to program files:} (to be added by Technical Editor) \\
{\em Developer's repository link:} https://github.com/plasmonX/plasmonX \\
{\em Licensing provisions:} GPLv3  \\
{\em Programming language:} Fortran 2008, Python           \\
{\em Nature of problem:} Simulating the response properties of plasmonic metallic and graphene-based nanomaterials.  \\
{\em Solution method:} Fully atomistic frequency-dependent fluctuating charges ($\omega$FQ) [1,2] and dipoles ($\omega$FQF$\mu$) [3] models and implicit, non-atomistic Boundary Element Methods (BEM) [4]. The approaches are implemented within the quasistatic approximation. \\
{\em Additional comments including restrictions and unusual features:}\\
  The program has been mainly tested by using gfortran (versions 9-13) combined with the Math Kernel Library (MKL) provided by Intel.
   \\

\end{small}
   \end{abstract}
\end{frontmatter}

\section{Introduction}

Plasmonic nanostructures, ranging from metal nanoparticles to graphene-based materials, have attracted significant interest due to their unique ability to confine light at the nanoscale, enabling applications ranging from enhanced spectroscopies and sensing to quantum optics and nanoelectronics \cite{maier2007plasmonics,stockman2011nanoplasmonics,gramotnev2010plasmonics,ishida2019importance,langer2019present,liz2006tailoring,grzelczak2008shape,giannini2011plasmonic,kelly2003optical}. The accurate simulation of their optical response, particularly in regimes where quantum effects such as tunneling and nonlocality become relevant \cite{scholl2012quantum,scholl2013observation,savage2012revealing,esteban2012bridging,esteban2015classical,baumberg2019extreme,esteban2012bridging,chen2019atomistic,teperik2013quantum,zhu2016quantum,urbieta2018atomic,savage2012revealing,marinica2012quantum,campos2019plasmonic,barbry2015atomistic}, remains a major theoretical and computational challenge.

Several numerical approaches have been developed to describe the response of plasmonic nanosystems \cite{zhao2008methods,jensen2008electronic,morton2011theoretical,corni2001enhanced,ringe2010unraveling,chen2015atomistic,jensen2009atomistic,chen2018morphology}. The most accurate methods rely on \textit{ab initio} quantum mechanical (QM) techniques, generally defined within the framework of time-dependent density functional theory (TDDFT) and its approximation (TD density functional tight binding -- TDDFTB) \cite{rossi2015quantized,rossi2020hot,kuisma2015localized,kumawat2025efficient,d2018density,chellam2025density,liu2022plasmon,herring2023recent,chaudhary2024optical,asadi2020td}. While these approaches allow for a rigorous treatment of the QM effects underlying the plasmonic response, their huge computational cost drastically limits the nanostructure sizes that can be simulated, which are generally composed of less than 1000 atoms \cite{chaudhary2024optical}, and are thus far from realistic, experimentally studied nanosystems. An attempt to overcome this limitation is offered by semiclassical and ``nonclassical'' electromagnetic models, which have been developed to incorporate key quantum corrections at a reduced computational cost \cite{hohenester2022nanoscale,mystilidis2023potential,ciraci2013hydrodynamic,ciraci2016quantum}. These include nonlocal, quantum hydrodynamic descriptions (and related extensions), quantum-corrected models that effectively account for tunneling in narrow gaps, and surface-response formulations based on Feibelman parameters \cite{yan2013green,toscano2015resonance,mystilidis2023opensans,eremin2022discrete,zheng2022boundary,li2017hybridizable,moeferdt2018plasmonic,huber2025computational}.

As an alternative, purely classical, non-atomistic approaches have been developed, such as the Discrete Dipole Approximation (DDA) \cite{draine1994discrete}, the Finite-Difference Time-Domain (FDTD) \cite{taflove2005computational}, and the Boundary Element Method (BEM) \cite{de2002retarded,myroshnychenko2008modeling,mennucci2019multiscale,bonatti2020frontiers,marcheselli2020simulating,coccia2020hybrid}, which model the nanostructure as a continuum dielectric characterized by its complex permittivity function. Such approaches are rooted in electrodynamics and solve Maxwell's equations by imposing the proper boundary conditions. Because they are based on classical physics, they can indeed afford large nanostructures; however, they lack any description of atomistic features and QM effects, and therefore, might fail to accurately describe 
regimes where quantum and mesoscopic effects become relevant \cite{mortensen2021mesoscopic,stamatopoulou2022finite,fitzgerald2016quantum}, including electron tunneling, nonlocal screening, Landau damping, quantum confinement, and atomic-scale defects, such as picocavities \cite{baumberg2022picocavities,benz2016single,giovannini2025electric}.

To bridge the gap between quantum accuracy and classical scalability, we have recently proposed novel atomistic, yet classical approaches for nanoplasmonics, named frequency-dependent fluctuating charges (\wfq)\cite{giovannini2019classical,giovannini2020graphene} and fluctuating dipoles (\wfqfmu) \cite{giovannini2022we}, which are specifically designed to account for the physical mechanisms underlying the optical response of plasmonic nanostructures. These methods offer a remarkable accuracy that is comparable to state-of-the-art ab initio TDDFT even for structures below the quantum size limit \cite{giovannini2022we,bonatti2020frontiers}, and are also able to account for non locality and quantum effects thanks to an effective description of quantum tunneling \cite{giovannini2019classical,giovannini2022we}, challenging the assumption that a full QM treatment is required to accurately model the plasmonic response. Also, being based on classical electrodynamics, \wfqfmu retains the positive aspects of classical approaches, i.e. it can be applied to nanostructure sizes that are untreatable at the ab initio level \cite{lafiosca2021going,giovannini2025electric}.

In this paper, we introduce \texttt{plasmonX}, a novel open-source code for simulating the plasmonic response of nanostructures, which features the first robust implementation of \wfq and \wfqfmu models. In addition, \texttt{plasmonX} provides a solid implementation of continuum dielectric methods (BEM), in a unified and efficient framework. Specifically, the dielectric polarizable continuum model (DPCM) \cite{tomasi2005quantum} and Integral Equation Formalism PCM (IEFPCM) \cite{tomasi1999ief,mennucci1997evaluation,cances1997new,corni2001enhanced} formalisms are implemented, supporting arbitrary geometries via mesh discretization. Differently from most simulation tools used in nanoplasmonics, which address plasmonic response either through continuum electrodynamics (e.g. MNPBEM \cite{trugler2012}), or through general-purpose quantum-mechanical methods, \texttt{plasmonX} is designed to bridge atomistic classical models and continuum approaches within a single, scalable software platform. In particular, \texttt{plasmonX} is the first open-source software that allows atomistic classical simulations of nanoplasmonics. The code is written in modern Fortran 2008, with a Python-based interface for input parsing, and a CMake build system that ensures cross-platform compatibility. It supports multiple numerical solvers, including matrix inversion, Krylov subspace iterative methods \cite{saad2003iterative,meurant2020krylov}, and a fully memory-efficient ``on-the-fly'' Generalized Minimum RESidual algorithm (GMRES) implementation \cite{saad1986gmres}, making it suitable for both small-scale and high-performance computing (HPC) applications for nanostructures of more than 1 million atoms \cite{lafiosca2021going}.

The manuscript is structured as follows: we briefly recall the theoretical foundations of the methods implemented in \texttt{plasmonX}, its modular architecture, and input specification, followed by performance benchmarks and illustrative simulations, including the optical properties of noble metal nanoalloys and field-enhancement in gold dimers. Finally, we introduce the \texttt{plasmonX} dedicated post-processing suite, which is distributed along with the main code, enabling visualization and analysis of induced densities and fields.

\section{Theoretical Models}
\label{Sec:Theory}

In this section, we provide a brief overview of the theoretical methods implemented in \texttt{plasmonX} ($\omega$FQ, $\omega$FQF$\mu$, and BEM), highlighting their common theoretical formalism, which allows for a modular implementation.

\subsection{\wfq}

$\omega$FQ \cite{giovannini2019classical,bonatti2020frontiers,giovannini2020graphene,bonatti2022silico}, endows each atom of the nanostructure with a complex charge that is a function of the frequency $\omega$ of the external perturbation. The charge exchange between the atoms is governed by the Drude conduction model, which describes the intraband contribution to plasmon decay. $\omega$FQ also features an effective description of quantum tunneling by means of a Fermi-like damping function, which introduces the typical exponential decay as a function of the atom-atom distance. We refer the interested reader to Ref. \citenum{giovannini2019classical} for the complete derivation of $\omega$FQ equations. 

The complex $\omega$FQ charges $q$ are obtained by solving the following complex linear system, written in a matrix form \cite{lafiosca2021going}:
\begin{equation}\label{eq:wfq-system}
\left[\overline{\mathbf{K}}\tqq-z_q(\omega)\mathbf{I}_N\right]\mathbf{q}(\omega) = -\overline{\mathbf{K}}\mathbf{V}(\omega)
\end{equation}
where $\overline{\mathbf{K}}$ is a $N \times N$ matrix whose elements read:
\begin{equation}
    \label{eq:kappa_bar}
    \overline{K}_{ij} = K_{ij} - \sum_k K_{ik}\delta_{ij}
\end{equation}
where $\delta_{ij}$ is the Kronecker delta. The elements of the $\mathbf{K}$ matrix take the following form:
\begin{equation}\label{eq:kappa}
K_{ij} = 
\begin{cases}
\frac{[1-f(r_{ij})]\mathcal{A}_i}{r_{ij}} & \mbox{if } i\ne j \\ 
0 & \mbox{if } i = j \\
\end{cases}
\end{equation}
$A_{i}$ is the effective area of atom $i$ through which the charge flux passes, $r_{ij}$ is the distance between the $i$-th and $j$-th atoms, while $f(r_{ij})$ is the Fermi-like damping function introduced to mimic quantum tunnelling:
\begin{equation}
    f(r_{ij}) = \frac{1}{1 + \exp\left[-d\left(\frac{r_{ij}}{s\cdot r_{ij}^0} - 1\right)\right]} 
    \label{eq:fermi}
\end{equation}
where $r^0_{ij}$ is the nearest neighbor distance, a fixed quantity specific to each lattice. In Eq. \ref{eq:wfq-system}, $\overline{\mathbf{K}}$ matrix is multiplied by $\mathbf{T}^{qq}$ matrix that takes into account the charge-charge interactions \cite{giovannini2019classical,mayer2007formulation}. In $\omega$FQ, each charge $q_i$, placed at $\mathbf{r_i}$, is associated with a Gaussian-type distribution of width equal to $R_{q_i}$: 
\begin{equation}
    \rho_{q_i}(\mathbf{r}) = \frac{q_i}{\pi^{3/2}R_{q_i}^3}\,\exp\left({-\frac{|\mathbf{r}-\mathbf{r}_i|^2}{R_{q_i}^2}}\right) 
\end{equation}
%
In this framework, the charge-charge interaction kernel $\mathbf{T}^{qq}$ reads \cite{mayer2007formulation}:
\begin{equation}\label{eq:tqq}
T_{ij}^{qq}=\frac{1}{|\mathbf{r}_{ij}|}\text{erf}\left(\frac{|\mathbf{r}_{ij}|}{R_{q_i-q_j}}\right)\\
\end{equation}
where $R_{q_i-q_j}=\sqrt{R_{q_i}^2+R_{q_j}^2}$. Note that the diagonal matrix of  $\mathbf{T}^{qq}$ is the chemical hardness. In Eq. \ref{eq:wfq-system}, $z_q(\omega)$ is a diagonal complex shift ($I^N$ is the $N \times N$ identity matrix) that accounts for the Drude mechanisms: 
\begin{equation}\label{eq:z_q_wfq}
    z_q(\omega) = -\frac{\omega}{2n\tau}(\omega\tau+\imm)
\end{equation}
where $n$ and $\tau$ are the static density and scattering time, respectively. $n$ can be expressed as:
\begin{equation}
    n = \frac{\sigma_0 m^{\star}}{\tau} = \frac{\omega_p^2 m^{\star}}{4\pi}
\end{equation}
where $\sigma_0$ is the static conductance of the nanomaterial, $\omega_p$ is the plasma frequency, and $m^\star$ is its effective electron mass. The latter can be approximated to 1 a.u. for metal nanoparticles \cite{pelton2013introduction}, while for doped graphene-based substrates, it can be expressed as: 
\begin{equation}\label{eq:2d-eff-mass}
m^\star = \frac{\sqrt{\pi n_{\mathrm{2D}}}}{v_F}
\end{equation}
where $n_\mathrm{2D}$ is the 2D electron density, and $v_F$ is the Fermi velocity \cite{neto2009electronic}, which is related to the Fermi energy $E_F$ through the expression $v_F = \sqrt{\frac{2E_F}{m_0}}$ ($m_0$ is the electron rest mass \cite{neto2009electronic}). The 3D electron density for graphene can be written in terms of $n_{\mathrm{2D}}$, which in turn depends on the value of $E_F$, which can be recovered directly from experimental conditions \cite{fang2013gated}.
Finally, in Eq. \ref{eq:wfq-system}, the right-hand side is made up of the product of the $\overline{\mathbf{K}}$ matrix and the external potential $\mathbf{V}(\omega)$ associated with the external field $\mathbf{E}(\omega)$.

\subsection{\wfqfmu}

$\omega$FQ is suitable for materials whose electronic properties can be well described by a Drude mechanism, such as simple metals (e.g., sodium) and graphene-based nanostructures. To describe $d-$metals (Ag, Au), the contribution of interband transitions must be included \cite{liebsch1993surface}. To this end, we introduce an additional polarization source, a dipole moment $\mu$, which, similar to the charge, is characterized by a Gaussian distribution of width equal to $R_{\mu_i}$ \cite{giovannini2022we}:
\begin{equation}
    \rho_{\mu_i} (\mathbf{r}) = \frac{|\bm{\mu}_i|}{\pi^{3/2}R_{\bm{\mu}_i}^3}\hat{n}_i\cdot \mathbf{\nabla}_{\mathbf{r}_i}\left[\exp\left({-\frac{|\mathbf{r}-\mathbf{r}_i|^2}{R_{\bm{\mu}_i}^2}}\right)\right]
\end{equation}
%
In the resulting \wfqfmu approach \cite{giovannini2022we}, the charges and dipole equations are solved simultaneously, yielding the following complex linear system:

\begin{equation}
\label{eq:wfqfmu-system}
\resizebox{\textwidth}{!}{$
\left[
\begin{pmatrix}
\overline{\mathbf{K}} & \mathbf{0} \\ 
\mathbf{0} & \mathbf{I}_{3N}
\end{pmatrix}
\begin{pmatrix}
\tqq & \tqmu \\ \tmuq & \tmumu
\end{pmatrix}
-\begin{pmatrix}
z_q(\omega)\mathbf{I}_N & \mathbf{0} \\ \mathbf{0} & z_\mu(\omega)\mathbf{I}_{3N}
\end{pmatrix}
\right]
\begin{pmatrix}
\mathbf{q}(\omega) \\ \bm{\mu}(\omega)
\end{pmatrix}
= 
\begin{pmatrix}
\overline{\mathbf{K}} & \mathbf{0} \\ 
\mathbf{0} & \mathbf{I}_{3N}
\end{pmatrix}
\begin{pmatrix}
-\mathbf{V}(\omega) \\ \mathbf{E}(\omega)
\end{pmatrix}
$}
\end{equation}
where $\mathbf{T}^{q\mu}$, and $\mathbf{T}^{\mu\mu}$ are the charge-dipole and dipole-dipole interaction kernels, which can be derived by proper differentiation of $\mathbf{T}^{qq}$ in Eq. \ref{eq:tqq}.\cite{mayer2007formulation,giovannini2019polarizable} In Eq. \ref{eq:wfqfmu-system}, $z_\mu(\omega)$ is defined as:
\begin{equation}\label{eq:z-mu}
z_\mu(\omega) = -\frac{1}{\alpha^\mathrm{IB}(\omega)}
\end{equation}
where $\alpha^\mathrm{IB}(\omega)$ is the interband polarizability that is extracted from the interband permittivity function, either calculated or experimentally measured \cite{giovannini2022we}.

Eq. \ref{eq:wfqfmu-system} is valid if all the system's atoms are of the same nature, i.e., for a homogeneous nanomaterial. To describe heterogeneous systems, i.e. characterized by two atom types, Eq. \ref{eq:wfqfmu-system} must be recast in the following form \cite{nicoli2023fully}: 

\begin{equation}\label{eq:wfqfmu-leghe}
\resizebox{\textwidth}{!}{$
\left[
\begin{pmatrix}
\frac{1}{2} (\overline{\mathbf{K}}+\overline{\mathbf{H}}(\omega))& \mathbf{0} \\ 
\mathbf{0} & \mathbf{I}_{3N}
\end{pmatrix}
\begin{pmatrix}
\tqq & \tqmu \\ 
\tmuq & \tmumu
\end{pmatrix}
-\begin{pmatrix}
z_q(\omega)\mathbf{I}_N & \mathbf{0} \\
\mathbf{0} & z_\mu(\omega)\mathbf{I}_{3N}
\end{pmatrix}
\right]
\begin{pmatrix}
\mathbf{q}(\omega) \\
\bm{\mu}(\omega)
\end{pmatrix}
= 
\begin{pmatrix}
\frac{1}{2} (\overline{\mathbf{K}}+\overline{\mathbf{H}}(\omega)) & \mathbf{0} \\ 
\mathbf{0} & \mathbf{I}_{3N}
\end{pmatrix}
\begin{pmatrix}
-\mathbf{V}(\omega) \\
\mathbf{E}(\omega)
\end{pmatrix}
$}
\end{equation}
where $z_{\mu,i}(\omega) = - 1/\alpha^{IB}_i(\omega)$ and:
\begin{equation}\label{eq:polar-finale}
\alpha^{IB}_i (\omega) \equiv \alpha^\mathrm{IB,hetero}_{A,i}(\omega) = \left(\frac{\frac{N_{A,i}+1}{\alpha^\mathrm{IB}_{A}(\omega)} + \frac{N_{B,i}}{\alpha^\mathrm{IB}_{B}(\omega)}}{N_{A,i}+N_{B,i}+1}\right)^{-1}
\end{equation}

\noindent where $N_{A,i}$ and $N_{B,i}$ are the number of nearest neighbors of $A$ and $B$ type, respectively. Furthermore, in Eq. \ref{eq:wfqfmu-leghe} the following quantities and matrices are introduced:
\begin{align}
\overline{H}_{ij}(\omega) & = H_{ij}(\omega) - \sum_{k} H_{ik}(\omega)\delta_{ij} \\
H_{ij}(\omega) & = \frac{n_j \tau_j}{n_i \tau_i}\frac{(1 - i \omega\tau_i)}{(1 - i\omega\tau_j)}K_{ji}
\end{align}
It is worth noting that, differently from \wfq and \wfqfmu, the off-diagonal elements of the left-hand side matrix in this case depend on the external frequency; however, this does not overall affect the general structure of the linear system. Furthermore, Eq. \ref{eq:wfqfmu-leghe} reduces to Eq. \ref{eq:wfqfmu-system} for a homogenous system.

\subsection{Boundary Element Method}

In BEM, the plasmonic nanostructure is represented as a continuous surface which is discretized into a mesh of triangular elements known as \textit{tesserae}. The key quantities computed in this approach are the surface charges, located at the barycenter of each \textit{tessera}. These surface charges determine the system’s optical response and depend on two factors: the geometry of the nanostructure and the complex, frequency-dependent permittivity of both the material and its surrounding medium.\cite{corni2001enhanced,tomasi2005quantum,Trugler}

In \texttt{plasmonX}, two BEM methods are implemented: the Dielectric Polarizable Continuum Model (DPCM) \cite{miertuvs1981electrostatic,Trugler,tomasi2005quantum} and the Integral Equation Formalism PCM (IEFPCM) \cite{cances1997new,tomasi2005quantum}. In the former, the BEM charges are calculated by solving the following linear system of equations:
\begin{equation}\label{eq:bem-dpcm}
    \bigg[ 2\pi \bigg(\frac{\varepsilon_2(\omega)+\varepsilon_1(\omega)}{\varepsilon_2(\omega)-\varepsilon_1(\omega)}\bigg)\textbf{A} + \textbf{F} \bigg]\bm{\sigma}(\omega) = -\textbf{E}_{\vec{n}}(\omega).
\end{equation}
while in IEFPCM, the charges are instead obtained as follows:
\begin{equation}\label{eq:bem-iefpcm}
    \bigg[ 2\pi \bigg(\frac{\varepsilon_2(\omega)+\varepsilon_1(\omega)}{\varepsilon_2(\omega)-\varepsilon_1(\omega)}\bigg) \mathbf{I}_N + {\textbf{F}}\bigg] \textbf{S}\textbf{A}^{-1}\bm{\sigma}(\omega) = - 
     \big(2\pi\mathbf{I}_T + \mathbf{F}\big)\mathbf{V}(\omega).
\end{equation}

In Eqs. \ref{eq:bem-dpcm} and \cref{eq:bem-iefpcm}, $\bm{\sigma}(\omega)$ are the discretized BEM surface charges, $\varepsilon_2(\omega)$ and $\varepsilon_1(\omega)$ are the frequency-dependent dielectric permittivities of the metal and the surrounding environment, respectively. The matrix \textbf{A} is diagonal and contains the area of each \textit{tessera}, and \textbf{I}$_N$ denotes the identity matrix for a system composed of $N$ \textit{tesserae}. In the case of BEM in DPCM formulation, $\textbf{E}_{\vec{n}}$ represents the component of the external electric field projected along the vector normal to each \textit{tessera}. In the IEFPCM formulation, $\textbf{V}$ corresponds to the electric potential evaluated at the barycenter of each \textit{tessera} \cite{cammi1995remarks,tomasi2005quantum}.

The \textbf{S} and \textbf{F} matrices depend on the geometric configuration of the mesh. The elements of the \textbf{S} matrix are given by \cite{cammi1995remarks,tomasi2005quantum}:
\begin{align*}
S_{ij} &=
\begin{cases}
\dfrac{A_j}{|\mathbf{s}_i - \mathbf{s}_j|}, & \text{if} \ i \neq j \\\\
1.0694 \sqrt{4\pi A_i}, & \text{if} \ i = j
\end{cases}
\end{align*}
\noindent where $A_i$, and $\mathbf{s}_i$ are the area and position of the $i$-th \textit{tessera}, respectively.

Two definitions of the \textbf{F} matrix, which contains the surface derivative of the Green function, are implemented in \texttt{plasmonX}. The first is an ``approximate'' method, valid for spherical and other simple-shaped geometries (e.g., rods), which evaluates the matrix elements as \cite{cammi1995remarks,tomasi2005quantum}:

\begin{align*}
F_{ij} &=
\begin{cases}
\dfrac{(\mathbf{s}_i - \mathbf{s}_j) \cdot \vec{\mathbf{n}}_j}{|\mathbf{s}_i - \mathbf{s}_j|^3} \cdot A_j, & \text{if} \ i \neq j \\\\
1.0694 \cdot \dfrac{\sqrt{4 \pi A_i}}{2R}, & \text{if} \ i = j
\end{cases}
\end{align*}

\noindent where $\vec{\mathbf{n}}_i$ is the normal vector of the $i$-th \textit{tessera}, and $R$ denotes the radius of the sphere enclosing the entire set of \textit{tesserae}, respectively. It is worth mentioning that some formulations instead determine the diagonal elements through linear combinations of the off-diagonal terms \cite{purisima1995simple}.

The second implementation involves a numerically more accurate evaluation of the surface derivative of the quasistatic Green function, which we refer to as the ``accurate'' method. In this case, the entries of the \textbf{F} matrix are computed by solving the following surface integral \cite{Trugler,trugler2012}:

\begin{equation}
    \int_{\Omega}  \frac{\partial G(\bm{s},\bm{s}')}{\partial n} \, d\Omega, \quad  G(\bm{s},\bm{s}') = \frac{1}{|\mathbf{s}-\mathbf{s}'|}
\end{equation}

\noindent where $G(\bm{s},\bm{s}')$ is the quasistatic Green function evaluated between surface points $\bm{s}$ and $\bm{s}'$. The symbol $\Omega$ denotes the surface of the nanoparticle, and $\partial n$  indicates differentiation with respect to the surface normal. For additional details on the ``exact'' integration procedure, the reader is referred to Refs. \citenum{Trugler} and \citenum{trugler2012}.

\section{Getting Started}

\texttt{plasmonX} is written in Fortran 2008 and is designed to be compiled and executed on multiple platforms with minimal user intervention. To facilitate its usage, the code architecture integrates a Python-based interface for input parsing and a CMake-based build system. The Python front-end is responsible for reading the simulation parameters from a human-readable YAML input file. This modular approach allows for a a clear separation between input specification and high-performance numerical execution in the Fortran backend.

The full source code is available on GitHub \cite{plasmonX_github} and makes use of two external components managed as git submodules. The pre-release version of \texttt{plasmonX} associated with this manuscript is archived on Zenodo \cite{plasmonx_zenodo}. To ensure a complete and functional installation, it is highly recommended to clone the repository using Git over SSH with the \verb|--recursive| option. Once the repository is cloned, a minimal Python environment must be configured. All required dependencies are listed in a \texttt{requirements.txt} or \texttt{requirements.yaml} file, depending on whether the user chooses \texttt{pip} or \texttt{conda} for package management. 

The compilation of the Fortran code is handled entirely through CMake~\cite{cmake}, with a custom Bash script (\texttt{setup.sh}) which is provided to streamline configuration across different operating systems. This script sets up the build directory, detects the available compilers, applies appropriate flags (e.g., OpenMP support), and runs the CMake configuration, with the default compiler flags and recommended build settings documented in the \texttt{plasmonX} user manual (see Ref.~\citenum{plasmonX_documentation}). The project is then compiled via standard \texttt{make} targets. The internal test suite is currently characterized by 83 tests, which ensure the correct compilation and linking with the external submodules. The tests are run by using \texttt{ctest}, which exploits the runtest package~\cite{runtest}, which is set as a submodule of the project. For each test case, the test suite includes a reference run log, generated on a Dell XPS 13, 11th Gen Intel(R) Core(TM) i7-1185G7 @ 3.00GHz, 32 GB RAM, which reports the wall-time at the end of the run, allowing users to compare results and sanity-check their installation.

The code supports Linux, macOS, and Windows environments (via Windows Subsystem for Linux)~\cite{wsl}. It has been tested with \texttt{gfortran} versions 9 through 13, and it is compatible with both standard LAPACK/BLAS libraries and the Intel Math Kernel Library (MKL)~\cite{mkl}, which is highly recommended for enhanced performance on Linux and Windows environments. The complete configuration and compilation workflow depends on the architecture and is amply documented within the user manual \cite{plasmonX_documentation}.

\section{Running and Input Configuration}

Once installed and tested, the \texttt{plasmonX} driver script can be executed by using the following command:

\begin{Verbatim}[commandchars=\\\{\}]
PATH/TO/BUILD/plasmonX.py [-h] -i input_file.yaml 
                          [-o output_file] 
                          [-omp omp_threads] 
                          [-mem memory_available]
\end{Verbatim}

The only required option is \texttt{-i}, which specifies the YAML input file. Optional arguments allow the user to set the output filename, number of OpenMP threads, and available memory for the calculation (see also Tab. \ref{tab:input_command}). The default output filename is \texttt{input\_file.log}, while the number of OpenMP threads and the available memory are set to the maximum available in the architecture exploited for the calculation. A help message can be accessed with the \texttt{-h} flag.

\begin{table}[!htbp]
   \begin{center}
   \begin{tabular}{l|c|p{8cm}}
   \hline
   \textbf{Option} & \textbf{Required} & \textbf{Description} \\
   \hline
   \texttt{-i} & Yes & Input file in YAML format \\
   \texttt{-o} & No & Output file name (default: \texttt{input\_file.log}) \\
   \texttt{-omp} & No & Number of OpenMP threads to use (default: all available) \\
   \texttt{-mem} & No & Available Memory in GB (default: system maximum) \\
   \texttt{-h} & No & Print help message and exit \\
   \hline
   \end{tabular}
   \end{center}
   \caption{List of the input options for running \texttt{plasmonX.py} script.}
   \label{tab:input_command}
\end{table}

The input to \texttt{plasmonX} is defined entirely by a YAML-formatted configuration file, typically named \texttt{input\_file.yaml}. The input is composed of various sections, each dedicated to a specific aspect of the simulation.

\begin{shaded}
\begin{Verbatim}[fontsize=\footnotesize]
what:
  # defines the type of calculation 
  [energy, static/dynamic response, restart]

forcefield:
  # defines the static/dynamic forcefield and the interaction kernels

field:
  # defines the field parameters [static/dynamic, frequencies, etc.]

algorithm:
  # defines the numerical algorithm to be used to solve response equations

output:
  # defines the parameters of the output printing

bem:
  # defines the parameters for implicit BEM calculations

control:
  # defines the parameters to control geometry/creation of additional files

atom_types:
  # defines the parameters of the atomtypes for atomistic simulations

parameters:
  # defines the parameters of the ωFQ and ωFQFμ models

input_geometry:
  # describes the atomic or implicit geometry in Ångstrom
\end{Verbatim}
\end{shaded}

This structure allows the user to define atomistic and continuum models in a unified format, while also keeping the input compact and readable. A number of default values (see \texttt{plasmonX} documentation \cite{plasmonX_documentation}) are automatically provided if certain sections are omitted, enabling simple and minimal configurations for the most user-friendly experience possible.To provide an overview of the \texttt{plasmonX} structure, Algorithm~\ref{alg:core_solver} summarizes the main driver workflow of the code. Further details on each section, including a complete list of available keywords and examples of \texttt{plasmonX} minimal input (i.e., exploiting default values), are provided in the documentation and examples (see Ref. \citenum{plasmonX_documentation}), and are discussed in the following.

\begin{algorithm}[!htbp]
\caption{Structure of \texttt{plasmonX}}
\label{alg:core_solver}
\begin{algorithmic}[1]
\State Read the input file
\State Build target object from input, through runtime polymorphism
\Statex \quad \texttt{target\_ $\leftarrow$ BEM / $\omega$FQ / $\omega$FQF$\mu$}
\If{Algorithm = inversion}
    \State Solve each frequency by direct inversion 
  \ElsIf{Algorithm = iterative}
    \If{Algorithm = iterative on the fly}
      \State Iterative solve with on-the-fly procedure 
    \Else
      \State Iterative solve with on-memory procedure
    \EndIf
  \EndIf
  \State Clean scratch; save data for post processing; deallocations
\end{algorithmic}
\end{algorithm}

\section{Specify the Nanostructure}

\texttt{plasmonX} potentially supports any kind of nanostructure by providing the atomic positions (for \wfq and \wfqfmu simulations) or the mesh tessellation (for BEM). Atomic positions are provided in the \texttt{input\_geometry} section, either by explicitly specifying the xyz coordinates by YAML inline string, or by specifying an XYZ file (case sensitive), which allows referencing external standard-format coordinates. The BEM mesh file is generally provided in the BEM section, under the keyword mesh. In this case, only msh files formatted in version \texttt{2.2 0 8} (ASCII and double precision) are allowed. \texttt{plasmonX} is agnostic to the specific material or symmetry of the nanostructure, and it is capable of handling arbitrary arrangements of atoms or continuum morphologies. The input geometry and the mesh are interpreted in units of \AA ngstrom. 

\begin{figure}[!htbp]
    \centering
    \includegraphics[width=0.9\linewidth]{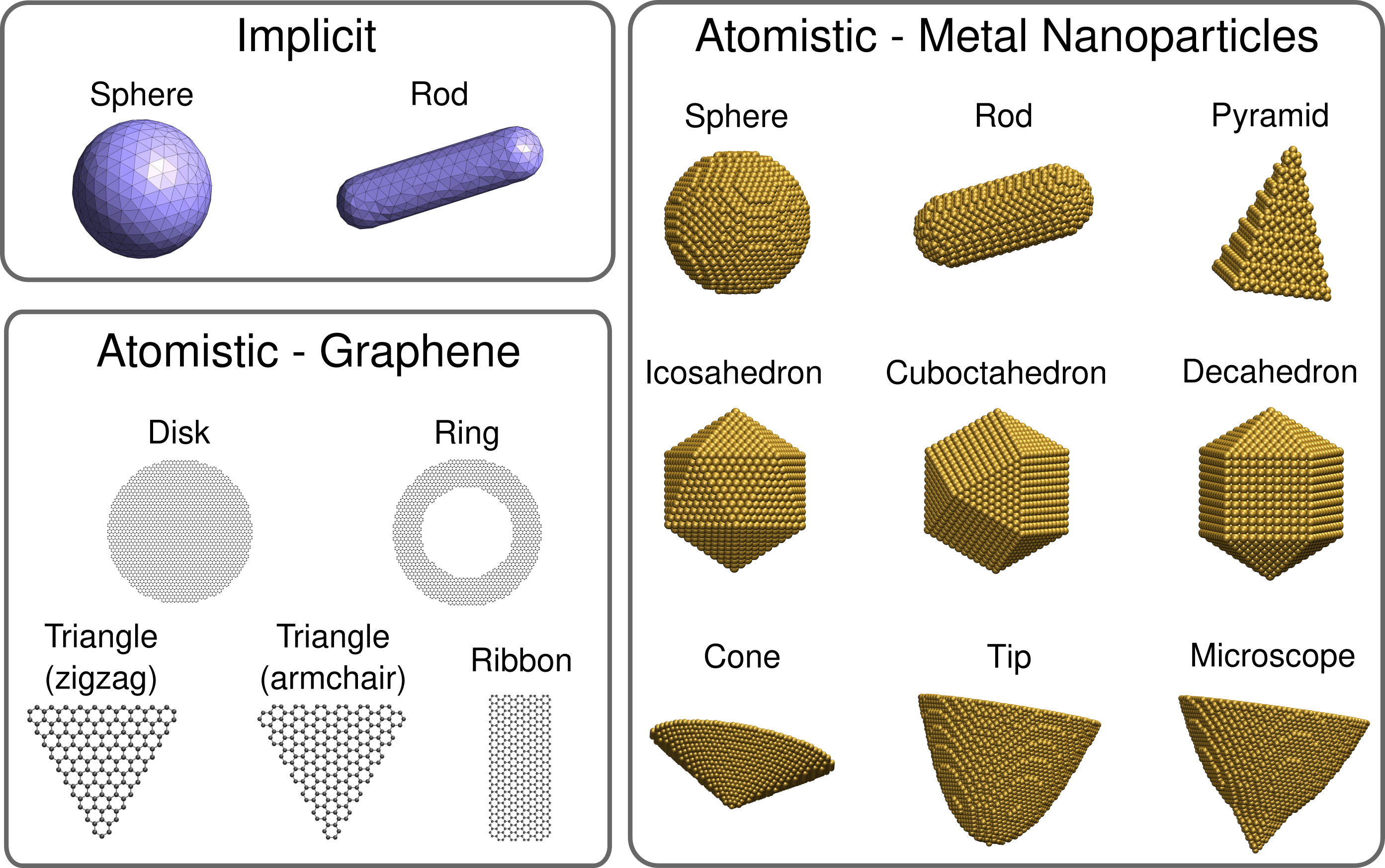}
    \caption{Nanostructure morphologies that can be created by \texttt{plasmonX} through the \texttt{GEOM} interface.}
    \label{fig:geom}
\end{figure}

\texttt{plasmonX} also provides a user-friendly input specification through the integrated interface to the \texttt{GEOM} library (commit: 1cc34d1 ~\cite{geom}). This interface enables the construction of both implicit and atomistic geometries by user specification of the relevant parameters, e.g., radius, shape, etc. For the implicit case, the allowed geometries are spheres and rods that are defined by specifying their type (e.g., \texttt{sphere}, \texttt{rod}) along with required parameters such as radius, length, orientation axis, and mesh resolution (we refer the interested user to the manual for further information). For atomistic geometries, \texttt{GEOM} supports a rich catalog of shapes, including spheres, rods, pyramids, icosahedra, cones, tips, and complex assemblies resembling experimental probe configurations (e.g., TERS tips). The supported morphologies through \texttt{GEOM} interface are graphically depicted in Fig. \ref{fig:geom}. The generator supports both homogeneous systems (single chemical composition) and, for some specific morphologies, heterogeneous systems (core-shell systems), with user-defined atom types for each region. Furthermore, carbon-based nanomaterials, such as graphene disks, rings, triangles, and ribbons, can be created with edge-type specification (armchair, zigzag). Again, we refer the interested user to the manual for details on the specific morphology \cite{plasmonX_documentation}. In all cases, the resulting geometries are passed to the Python-based preprocessor, which parses the YAML input and creates the specific geometry (atomistic or mesh) through the \texttt{GEOM} interface, for the subsequent computation in the Fortran backend.

\section{Solvers}

All the methods implemented in \texttt{plasmonX} are characterized by a similar formal structure \cite{rodriguez2025quantum}. In fact, in all cases, the response variables (charges and dipoles) are obtained by following a complex linear system that can be written as: 

\begin{equation}
    \left[ \mathbf{M}(\omega) - \mathbf{Z}(\omega) \right] \mathbf{L}(\omega) = \mathbf{U}(\omega) \mathbf{R}(\omega)
    \label{eq:matrix-compact}
\end{equation}

The definition of all matrices and vectors in Eq. \ref{eq:matrix-compact} can be recovered by comparing with Eq. \ref{eq:wfq-system} (\wfq), Eq. \ref{eq:wfqfmu-system} (\wfqfmu for homogeneous systems), Eq. \ref{eq:wfqfmu-leghe} (\wfqfmu for heterogeneous systems), Eq. \ref{eq:bem-dpcm} (BEM in DPCM formulation), and Eq. \ref{eq:bem-iefpcm} (BEM in IEFPCM formulation). 
We note that for \wfqfmu for heterogeneous systems (Eq. \ref{eq:wfqfmu-leghe}) the $\mathbf{M}$ and $\mathbf{U}$ matrices are frequency dependent and complex, while they are static and real for all other methods. $\mathbf{Z}(\omega)$ and $\mathbf{L}(\omega)$ are frequency-dependent and complex in all cases, while $\mathbf{R}(\omega)$ is always real and contains the external polarization sources, such as the external potential and field. 

To solve Eq. \ref{eq:matrix-compact}, \texttt{plasmonX} implements three algorithms, which are common in this kind of simulations \cite{lipparini2014scalable,lafiosca2021going,bugeanu2015wavelet,hohenester2018making}: 
\begin{enumerate}
    \item Inversion: the complex linear system is solved by resorting to LU factorization and subsequent resolution of the complex linear systems through an interface to LAPACK subroutines (ZGETRF and ZGETRS). This is the default for small systems. 
    \item Iterative: the complex linear system is solved by resorting to the iterative Krylov GMRES method \cite{saad1986gmres}. \texttt{plasmonX} provides an effective implementation of complex GMRES for all methods by storing the matrices in memory \cite{lafiosca2021going}.
    \item Iterative on the fly: the complex linear system is solved by resorting to GMRES \cite{saad1986gmres}, and the matrix product multiplications are performed on the fly, without storing the matrices in memory \cite{lafiosca2021going}. This is implemented for \wfq and \wfqfmu methods only, and is the default for large systems.
\end{enumerate}

\begin{figure}[!htbp]
    \centering
    \includegraphics[width=.45\linewidth]{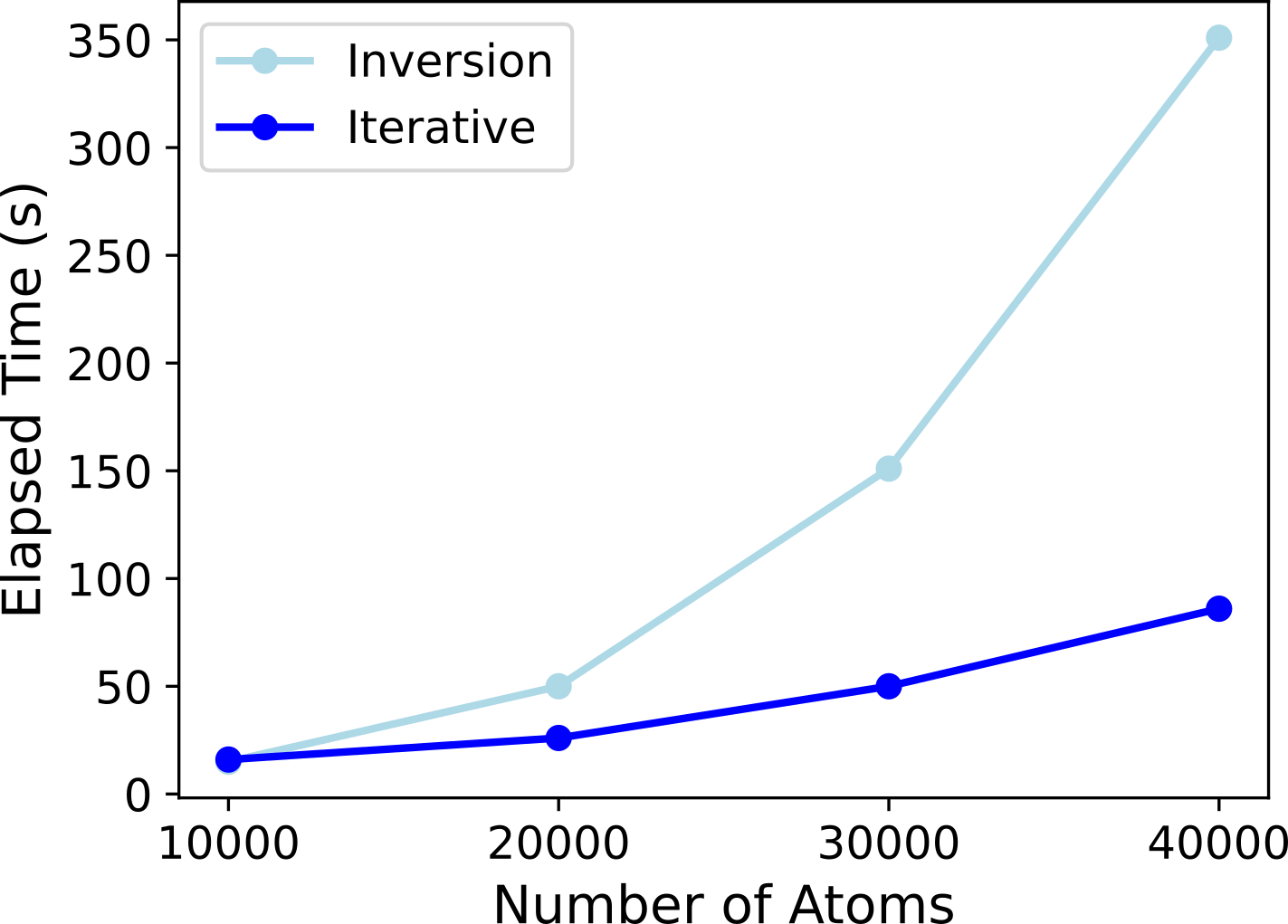}
    \caption{Elapsed time required to solve the \wfq linear system as a function of the number of atoms for selected graphene disks at the plasmon resonance frequency by using inversion and iterative (on the fly) algorithms implemented in \texttt{plasmonX}.}
    \label{fig:gfortran_n_atoms}
\end{figure}

For both iterative solvers, the numerical convergence is controlled by the residual norm of the linear system by means of the root-mean-square error (RMSE). A default tolerance of $|E_0| \cdot 10^{-5}$ is employed (where $|E_0|$ is the applied electric field intensity), which we have previously shown to provide a reliable compromise between numerical accuracy and computational cost \cite{lafiosca2021going}. In Fig. \ref{fig:gfortran_n_atoms}, we provide a graphical depiction of the elapsed time (in seconds) of inversion and iterative (on-the-fly) algorithms as a function of the number of atoms of graphene disks with a radius ranging from 10 nm to 19 nm (from 10k to 40k atoms). The calculations are performed at the plasmon resonance frequency of each disk (10k atoms: 0.30 eV; 20k atoms: 0.26 eV; 30k atoms: 0.24 eV; 40k atoms: 0.22 eV) by imposing a Fermi Energy of 0.4 eV. For the iterative solution, the Root Mean Squared Error (RMSE) is set to $10^{-8}$ for an applied electric field with $E_0 = 10^{-4}$ a.u., which we have previously shown to provide the best compromise between computational cost and accuracy \cite{lafiosca2021going}. All calculations are performed on a machine equipped with an Intel(R) Xeon(R) Gold 6252N CPU @ 2.30GHz, 96 threads, and 512 GB of RAM. 
As can be seen for the smallest system ($\sim$ 10k atoms), the inversion and iterative algorithms require almost the same elapsed time (16 s vs. 15 s). By increasing the size, the inversion algorithm follows a third power scaling ($y \propto P(x^3), R^2 > 0.99$), while the iterative one follows a second power progression ($y \propto P(x^2), R^2 > 0.99$), rapidly becoming the most efficient algorithm to solve the linear complex problem. For a detailed analysis of the performance of the GMRES algorithm in solving the \wfq linear system, we refer the interested reader to Ref. \citenum{lafiosca2021going}.

\begin{figure}[!htbp]
    \centering
        \includegraphics[width=0.5\linewidth]{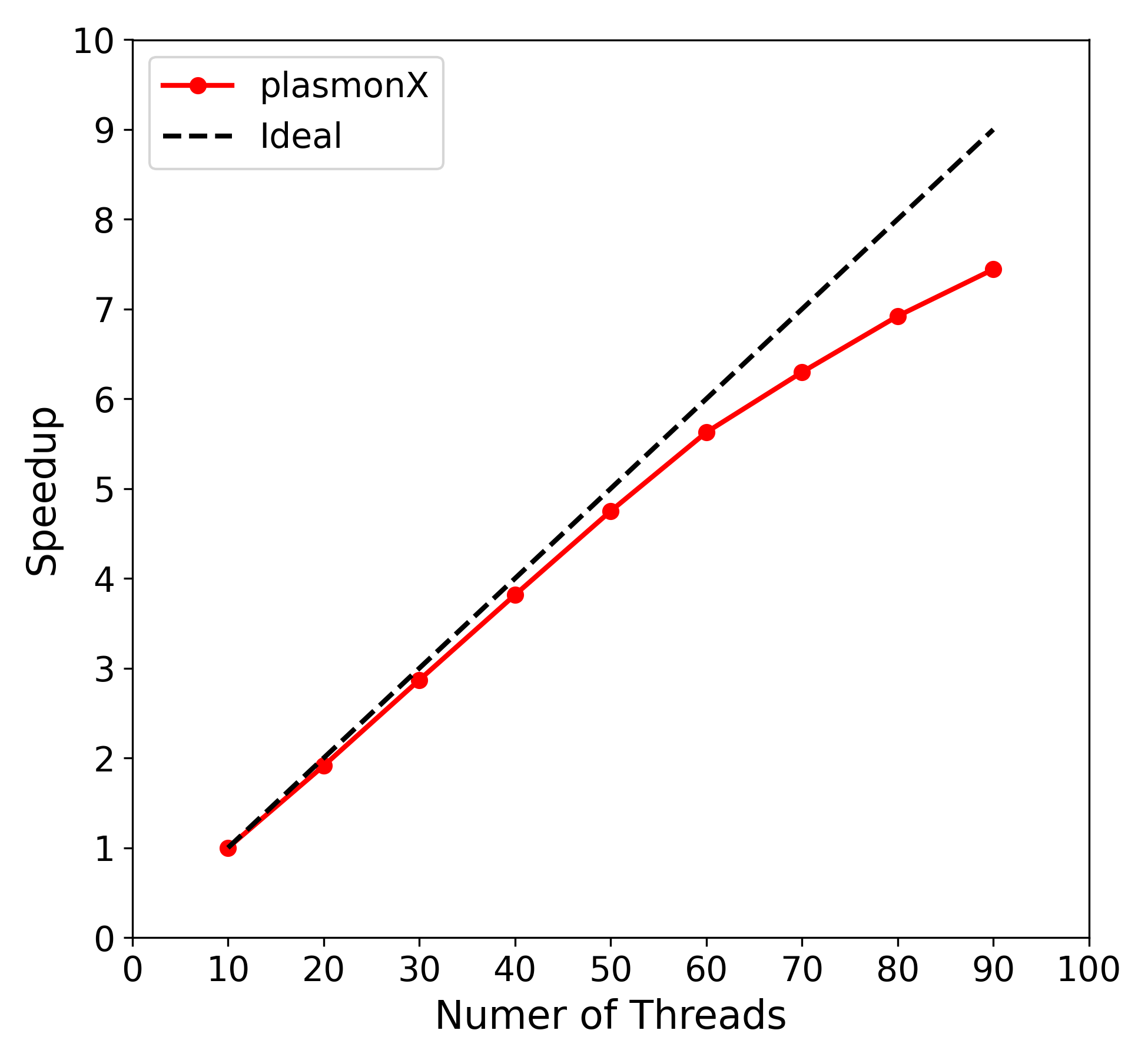}
    \caption{Speedup of the \wfqfmu linear system solution using the iterative on-the-fly algorithm as a function of the OMP threads.}
    \label{fig:speedup}
\end{figure}
The iterative on-the-fly algorithm is especially useful for large systems because the requirements in terms of RAM are particularly low, as only a few vectors are stored in memory during the run. To showcase the performance of \texttt{plasmonX} implementation, we perform 9 calculations by increasing the number of processors from 10 to 90 with steps of 10. Calculations employ the \wfqfmu approach and are run on a Silver nanoparticle with icosahedral shape with a radius of about 10 nm, i.e. composed of about 149k atoms (about 600k variables). We then compute the speedup of the calculation as follows: 
\begin{equation}
    \mathrm{speedup}(n) = \dfrac{T(n)}{T(10)}
\end{equation}
where $T(n)$ is the elapsed time exploiting $n$ OMP threads, while $T(10)$ is the reference obtained by using 10 OMP threads. Ideally, by moving from 10 to 90 OMP threads, the maximum speedup is 9. The computed speedups are graphically depicted in Fig. \ref{fig:speedup}. The code demonstrates a nearly linear speedup up to 60 threads, with parallel efficiency (calculated as $\mathrm{speedup}(n)/(n/10)$) remaining above 90\%. Beyond this point, the efficiency gradually decreases, reaching approximately 84\% at 90 threads. This decrease is consistent with the presence of parallel overheads and resource contention effects, which are typical in simulations of this kind \cite{hager2010hpc}. Nevertheless, the observed speedup of 7.58 at 90 threads (relative to 10 threads) indicates that the code maintains a high degree of parallel effectiveness, which can be exploited in HPC applications.

\section{Observables}

Once the complex linear system defining each model is been solved for a specific $\omega$, the total complex dipole moment $\overline{\mathbf{d}}(\omega)$ of the system can be calculated as \cite{giovannini2019classical,giovannini2022we}:
\begin{align}
    \omega\text{FQ} : \quad \overline{\mathbf{d}}(\omega) & = \sum_i q_i(\omega) \mathbf{r}_i \\
    \omega\text{FQF}\mu : \quad \overline{\mathbf{d}}(\omega) & = \sum_i q_i(\omega) \mathbf{r}_i + \bm{\mu}_i (\omega) \\
    \text{BEM} : \quad \overline{\mathbf{d}}(\omega) & = \sum_i \sigma_i(\omega) \mathbf{s}_i 
\end{align}
from which the complex polarizability $\overline{\alpha}$ is obtained:
\begin{equation}
    \overline{\alpha}_{kl} (\omega) = \frac{\partial \overline{d}_k (\omega) }{\partial E_l(\omega)}  \qquad k,l \in \{x,y,z\}
\end{equation}
From $\overline{\alpha}$, absorption, scattering, and extinction cross sections can finally be evaluated as:
\begin{align}
    \sigma^{\text{abs}} (\omega) & = \frac{4\pi\omega}{3c} \text{Tr}\ \{ \text{Im} \ \overline{\alpha}(\omega)\} \\ 
    \sigma^{\text{sca}} (\omega) & = \frac{8\pi\omega^4}{3c^4} \ |\overline{\alpha}(\omega)|^2\\
    \sigma^{\text{ext}} (\omega) & = \sigma^{\text{abs}} (\omega) + \sigma^{\text{sca}} (\omega)
\end{align}
where $c$ is the speed of light. Since \texttt{plasmonX} implements methods in the quasi-static regime, the scattering cross section is generally negligible. 

\begin{figure}[!htbp]
    \centering
    \includegraphics[width=0.7\linewidth]{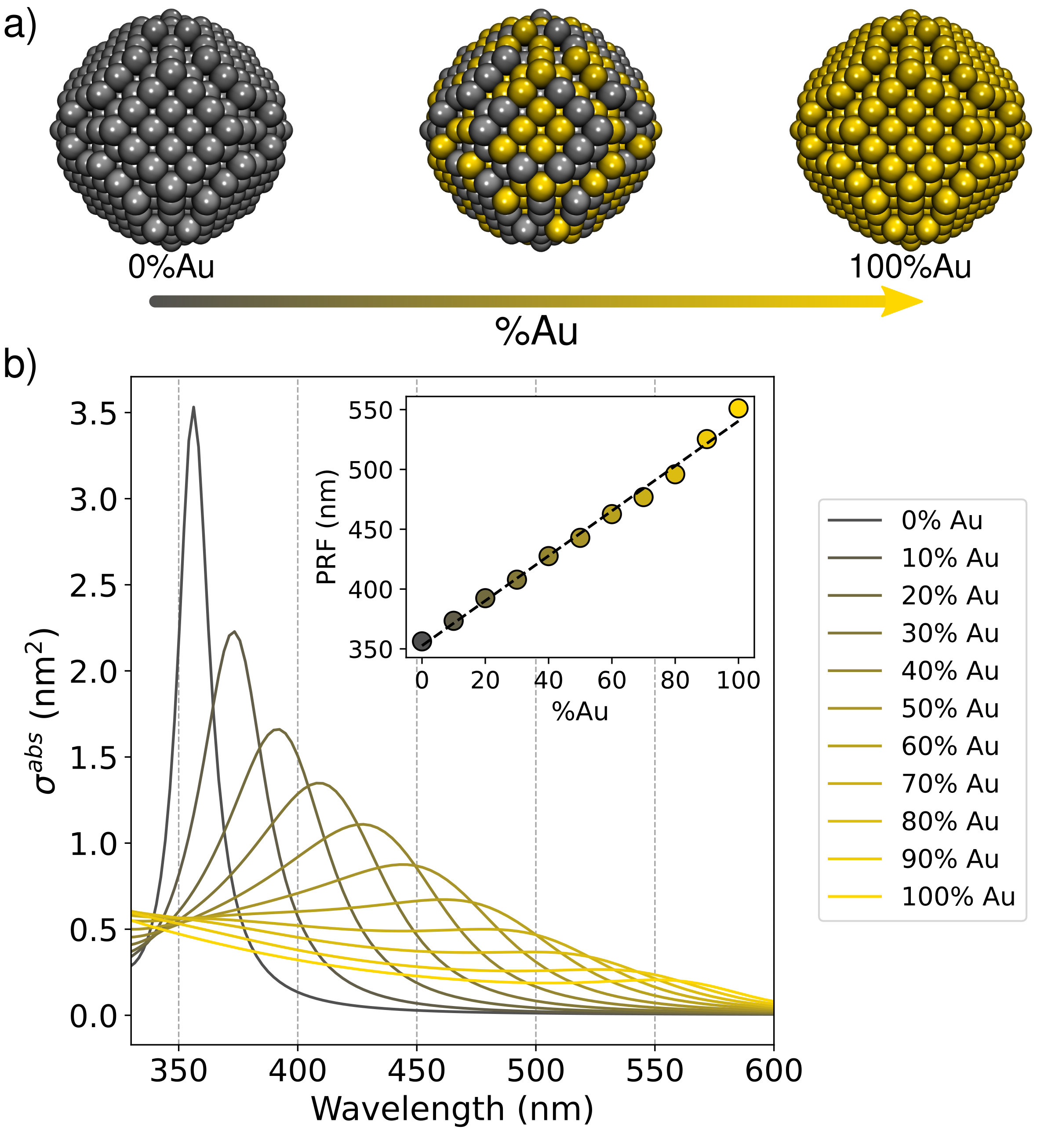}
    \caption{(a) Graphical depiction of spherical Ag-Au nanoalloys. (b) \wfqfmu plasmon resonance frequency (PRF) as a function of Au percentage.}
    \label{fig:alloys}
\end{figure}

To showcase the results that can be obtained with \texttt{plasmonX} we consider ten Ag-Au spherical nanoalloys with diameter D = 3.5 nm (1289 atoms) which are constructed by increasing the percentage of Au atoms in a Ag spherical nanoparticle created by using GEOM interface (see Fig. \ref{fig:alloys}a), randomly replacing Ag atoms (from 0\% to 100\%, with a constant step of 10\%). Once the structures are created, the \wfqfmu calculation is performed by running the following minimal input:
\begin{shaded}
\begin{Verbatim}[fontsize=\footnotesize]
what:
  - dynamic response

forcefield:
  dynamic: wFQFMu

field:
  min freq: 2.0
  max freq: 4.5
  step freq: 0.01

input_geometry:
  external xyz file: Au_percentage.xyz
\end{Verbatim}
\end{shaded}
where \texttt{percentage} corresponds to the Au percentage of each nanoalloy. The Au-Ag parameters are set to the default values and are taken from Refs. \citenum{giovannini2022we} and \citenum{nicoli2023fully}.

The computed absorption cross sections $\sigma^{\text{abs}}$ for all the studied alloys are reported in Fig. \ref{fig:alloys}b.
All spectra are characterized by a main peak, which redshifts and lowers in intensity as the Au concentration increases. By plotting the PRF (in nm) as a function of the percentage of Au (see Fig. \ref{fig:alloys}b), a linear trend is observed ($R^2 > 0.99$, see inset). This is in agreement with Vegard's law, which is reported by various experiments \cite{link1999alloy,rioux2014analytic,papavassiliou1976surface,ristig2015nanostructure}:
\begin{equation}
    \label{eq:vegard}
    \lambda^{Vegard}(x) = (1-x) \lambda_{Ag} + x \lambda_{Au} 
\end{equation}
where $x$ represents the Au percentage. The linear slope (1.88 nm/\% Au) of the fitted line is also in very good agreement with experiments reported in Ref. \citenum{link1999alloy} ($\sim$ 1.35 nm/\% Au). Also, the computed intensity of the plasmon band exponentially decreases by increasing the Au percentage, perfectly reproducing experimental findings \cite{link1999alloy}. The excellent agreement provided by \wfqfmu is remarkable, especially considering that other classical approaches fail at reproducing the experimentally reported linear trend in wavelength and the exponentially decreasing intensity \cite{sorensen2021atomistic}. 
We conclude this section by highlighting that the considered atomistic structures are intentionally small (1289 atoms) to demonstrate that physically relevant results, that can be directly compared with experiments, can be obtained by running \texttt{plasmonX} on a personal laptop (in this case, Dell XPS 13" equipped with 11th Gen Intel(R) Core(TM) i7-1185G7 @ 3.00GHz and 32GB of RAM).

\section{Post-process Analysis}

By default, \texttt{plasmonX} creates a tar.gz archive where the most relevant information of the calculation is stored, including that for post-process analysis. \texttt{plasmonX} is distributed together with a post-process code, named \texttt{plasmonX\_analysis}, that gives access to various analyses, such as the extraction of XYZ and PQR files (for \wfq and BEM calculations), CUBE and PLT files for plotting induced densities and fields, and CSV files for the analysis of the induced densities and fields in user-defined planes. The code is compiled together with the main code \texttt{plasmonX}, and uses the modularity and object-oriented implementation to access the specificities of each model. As for the main code, the user-provided input is managed by a Python interface. All options are presented and discussed in the user manual.

\begin{figure}[!htbp]
    \centering
    \includegraphics[width=0.7\linewidth]{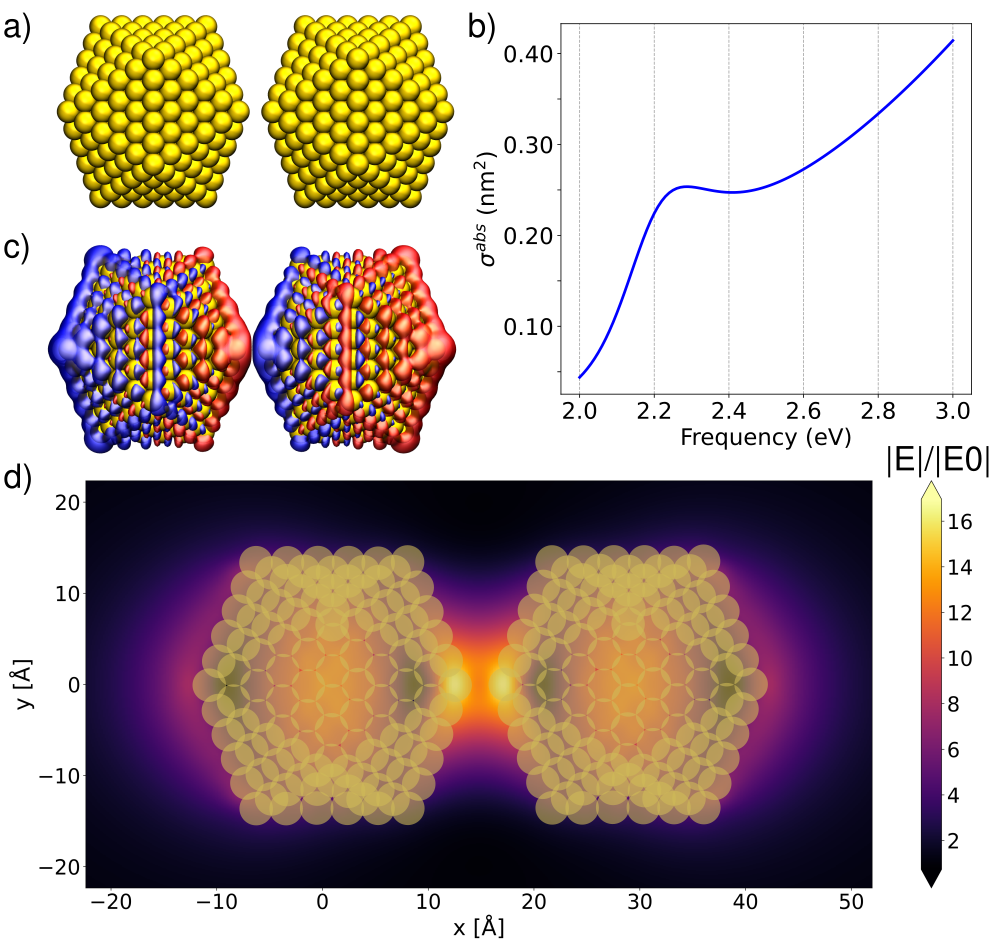}
    \caption{(a) Graphical depiction of Au$_{561}$ dimer; (b-d) \wfqfmu absorption cross section (b), induced density (c), and enhanced electric field computed at the PRF (2.28 eV) using a polarization field along the $x$ axis.}
    \label{fig:Au_dimer}
\end{figure}

To showcase the performance of \texttt{plasmonX\_analysis}, we select an atomistic Gold Ih dimer (gap separation along the $x$ axis of 5 \AA), constructed by using the GEOM interface and described at the \wfqfmu level (see Fig. \ref{fig:Au_dimer}a). The resulting dimer is composed of 1122 atoms. We highlight that \wfqfmu can accurately describe similar systems characterized by a subnanometer gap \cite{giovannini2019classical,zhu2016quantum,urbieta2018atomic,marinica2012quantum,esteban2015classical,duan2012nanoplasmonics,scholl2012quantum,scholl2013observation}, as it physically accounts for quantum tunneling effects (see Eq. \ref{eq:fermi}) \cite{giovannini2019classical}. The \wfqfmu calculation is performed by using the following minimal input, which exploits the default Au \wfqfmu parameters taken from Ref. \citenum{giovannini2022we}:
\begin{shaded}
\begin{Verbatim}[fontsize=\footnotesize]
what:
  - dynamic response

forcefield:
  dynamic: wFQFMu

field:
  min freq: 2.0
  max freq: 3.0
  step freq: 0.01

input_geometry:
  shape: Ih
  atomtype: Au
  radius: 15.0
  distance: 5.0
  dimer_axis: +x
\end{Verbatim}
\end{shaded}

The resulting absorption cross section is graphically depicted in Fig. \ref{fig:Au_dimer}b, and clearly shows a main plasmon peak at 2.28 eV. By using \texttt{plasmonX\_analysis}, the nature of the plasmon peak can be analyzed by calculating the density induced at the PRF by an external field polarized along the $x$ axis. The resulting induced density is graphically depicted in Fig. \ref{fig:Au_dimer}c, and is obtained by running the following command:
\begin{shaded}
\begin{Verbatim}[commandchars=\\\{\},fontsize=\footnotesize]
PATH/TO/BUILD/plasmonX_analysis.py -i output_file.tar -w density -n 1 \\
                                   -freq 2.28 -field_dir x
\end{Verbatim}
\end{shaded}
which creates a CUBE file that can be used to graphically plot the induced density. The graphical rendering can be performed by using an external visualization tool (e.g. VMD\cite{vmd}, as in the present case). Fig. \ref{fig:Au_dimer}c shows that the plasmon peak is associated with a boundary dipolar plasmon (BDP), i.e. the two plates are characterized by an induced plasmon with dipolar character in the direction of the polarization field. As stated above, \texttt{plasmonX\_analysis} also gives access to the calculation of the induced field, which is a fundamental quantity for surface-enhanced spectroscopies \cite{lafiosca2023qm,langer2019present,giovannini2025modeling,illobre2025mixed}. In this case, we compute the electric field in the $xy$ plane as induced by the same external field as above, by using the following command:
\begin{shaded}
\begin{Verbatim}[commandchars=\\\{\},fontsize=\footnotesize]
PATH/TO/BUILD/plasmonX_analysis_py -i output_file.tar.gz -w field -n 1 \\
                                   -freq 2.28 -field_dir x -plane xy \\
                                   -n_plane 1
\end{Verbatim}
\end{shaded}
The resulting 2D map is graphically depicted in Fig. \ref{fig:Au_dimer}d. This is achieved by utilizing the Python Matplotlib package; remarkably, \texttt{plasmonX\_analysis} provides the user with a Python script that can be modified for customization. Fig. \ref{fig:Au_dimer}c shows that at the PRF, the studied dimer sustains a highly localized hot spot in the middle of the gap, correctly reproducing ab initio results for similar systems. 

\section{Conclusions and Outlook}

In this work, we have presented \texttt{plasmonX}, a novel open-source platform for simulating the linear response of plasmonic nanostructures. The code features a unified implementation of both atomistic and continuum electrodynamics models, $\omega$FQ and $\omega$FQF$\mu$ and BEM in both DPCM and IEFPCM formalisms. Thanks to its modular architecture, flexible input design, and efficient numerical solvers, which include iterative and memory-efficient “on-the-fly” implementations, \texttt{plasmonX} enables accurate and scalable simulations from hundreds to millions of atoms. 

Several key developments are planned to further expand the scope and efficiency of \texttt{plasmonX}. First, we aim to extend the current frequency-domain formalism to a real-time (RT) framework, allowing the simulation of ultrafast plasmonic phenomena, by implementing our recently developed RT-\wfq and RT-\wfqfmu methods \cite{lafiosca2024real}. Second, we are actively working on generalizing the models to describe condensed phase systems, such as colloids and nanostructures dissolved in an external environment, by implementing the recently developed multiscale framework which integrates \wfq and \wfqfmu with polarizable embedding \cite{nicoli2024atomistic}. 

A further avenue involves the development of mixed implicit/explicit strategies, in which part of the nanostructure is described using a full atomistic treatment (for instance, the surface) while the remaining (the core) is treated at the BEM level. We have recently proved that such partitioning drastically reduces the computational cost without reducing the overall accuracy of the \wfq/\wfqfmu methods \cite{illobre2025mixed}. Finally, to extend the applicability of \texttt{plasmonX} to large and extended nanostructures, we plan to incorporate retardation effects, enabling the description of plasmonic phenomena beyond the quasistatic regime. To this end, we aim to implement fast multipole algorithms \cite{darve2000fast}, which would enable linear or near-linear scaling of the long-range electrostatic interactions, making \texttt{plasmonX} suitable for simulations involving several millions of atoms.

\section*{Data availability}

The data supporting this study are provided in the paper. The full simulation data discussed in Sections 7 and 8 are available upon reasonable request.

\section*{CRediT author statement}

\textbf{Tommaso Giovannini}: Conceptualization, Methodology, Software, Validation, Investigation, Writing - Original Draft, Writing - Review \& Editing, Visualization, Supervision, Project administration. \textbf{Pablo Grobas Illobre}: Methodology, Software, Validation, Investigation, Writing - Original Draft, Visualization. \textbf{Piero Lafiosca}: Methodology, Software, Validation. \textbf{Luca Nicoli}:  Methodology, Software, Validation. \textbf{Luca Bonatti}: Methodology, Software, Validation. \textbf{Stefano Corni}: Conceptualization,  Methodology, Writing - Review \& Editing, Supervision. \textbf{Chiara Cappelli}: Conceptualization, Methodology, Resources, Writing - Review \& Editing, Supervision, Funding acquisition, Project administration.

\section*{Declaration of Competing Interest}

The authors declare that they have no known competing financial interests or personal relationships that could have appeared to influence the work reported in this paper.

\section*{Acknowledgments}

This work has received funding from the European Research Council (ERC) under the European Union’s Horizon 2020 research and innovation programme (Grant Agreement No. 818064). We gratefully acknowledge the Center for High-Performance Computing (CHPC) at SNS for providing the computational infrastructure.  CC and PGI acknowledge funding from MUR-FARE Ricerca in Italia: Framework per l’attrazione ed il rafforzamento delle eccellenze per la Ricerca in Italia - III edizione. Prot. R20YTA2BKZ. TG acknowledges financial support from the University of Rome Tor Vergata ``Ricerca Scientifica di Ateneo 2024'' INNOVATIONS. SC acknowledges funding from the European Union under the  grant ERC-2015-CoG-681285.






\begin{thebibliography}{0}
\bibitem{1} T. Giovannini, M. Rosa, S. Corni, C. Cappelli, A classical picture of subnanometer junctions: an atomistic Drude approach to nanoplasmonics, Nanoscale 11 (13) (2019) 6004–6015         
\bibitem{2} T. Giovannini, L. Bonatti, M. Polini, C. Cappelli, Graphene plasmonics: Fully atomistic approach for realistic structures, J. Phys. Chem. Lett. 11 (18) (2020) 7595–7602.         
\bibitem{3} T. Giovannini, L. Bonatti, P. Lafiosca, L. Nicoli, M. Castagnola, P. G. Illobre, S. Corni, C. Cappelli, Do we really need quantum mechanics to describe plasmonic properties of metal nanostructures?, ACS Photonics 9 (9) (2022) 3025–3034.
\bibitem{4} F. J. García de Abajo, A. Howie, Retarded field calculation of electron energy loss in inhomogeneous dielectrics, Phys. Rev. B 65 (11) (2002) 115418.
\end{thebibliography}

\begin{thebibliography}{100}
\expandafter\ifx\csname url\endcsname\relax
  \def\url#1{\texttt{#1}}\fi
\expandafter\ifx\csname urlprefix\endcsname\relax\def\urlprefix{URL }\fi
\expandafter\ifx\csname href\endcsname\relax
  \def\href#1#2{#2} \def\path#1{#1}\fi

\bibitem{maier2007plasmonics}
S.~A. Maier, Plasmonics: fundamentals and applications, Springer Science \& Business Media, 2007.

\bibitem{stockman2011nanoplasmonics}
M.~I. Stockman, Nanoplasmonics: past, present, and glimpse into future, Opt. Expr. 19~(22) (2011) 22029--22106.

\bibitem{gramotnev2010plasmonics}
D.~K. Gramotnev, S.~I. Bozhevolnyi, Plasmonics beyond the diffraction limit, Nat. Photon. 4~(2) (2010) 83--91.

\bibitem{ishida2019importance}
T.~Ishida, T.~Murayama, A.~Taketoshi, M.~Haruta, Importance of size and contact structure of gold nanoparticles for the genesis of unique catalytic processes, Chem. Rev. 120~(2) (2019) 464--525.

\bibitem{langer2019present}
J.~Langer, D.~Jimenez~de Aberasturi, J.~Aizpurua, R.~A. Alvarez-Puebla, B.~Auguié, J.~J. Baumberg, G.~C. Bazan, S.~E.~J. Bell, A.~Boisen, A.~G. Brolo, J.~Choo, D.~Cialla-May, V.~Deckert, L.~Fabris, K.~Faulds, F.~J. García~de Abajo, R.~Goodacre, D.~Graham, A.~J. Haes, C.~L. Haynes, C.~Huck, T.~Itoh, M.~Käll, J.~Kneipp, N.~A. Kotov, H.~Kuang, E.~C. Le~Ru, H.~K. Lee, J.-F. Li, X.~Y. Ling, S.~A. Maier, T.~Mayerhöfer, M.~Moskovits, K.~Murakoshi, J.-M. Nam, S.~Nie, Y.~Ozaki, I.~Pastoriza-Santos, J.~Perez-Juste, J.~Popp, A.~Pucci, S.~Reich, B.~Ren, G.~C. Schatz, T.~Shegai, S.~Schlücker, L.-L. Tay, K.~G. Thomas, Z.-Q. Tian, R.~P. Van~Duyne, T.~Vo-Dinh, Y.~Wang, K.~A. Willets, C.~Xu, H.~Xu, Y.~Xu, Y.~S. Yamamoto, B.~Zhao, L.~M. Liz-Marzán, Present and future of surface-enhanced raman scattering, ACS Nano 14~(1) (2020) 28--117.

\bibitem{liz2006tailoring}
L.~M. Liz-Marz{\'a}n, Tailoring surface plasmons through the morphology and assembly of metal nanoparticles, Langmuir 22~(1) (2006) 32--41.

\bibitem{grzelczak2008shape}
M.~Grzelczak, J.~P{\'e}rez-Juste, P.~Mulvaney, L.~M. Liz-Marz{\'a}n, Shape control in gold nanoparticle synthesis, Chem. Soc. Rev. 37~(9) (2008) 1783--1791.

\bibitem{giannini2011plasmonic}
V.~Giannini, A.~I. Fern{\'a}ndez-Dom{\'\i}nguez, S.~C. Heck, S.~A. Maier, Plasmonic nanoantennas: fundamentals and their use in controlling the radiative properties of nanoemitters, Chem. Rev. 111~(6) (2011) 3888--3912.

\bibitem{kelly2003optical}
K.~L. Kelly, E.~Coronado, L.~L. Zhao, G.~C. Schatz, The optical properties of metal nanoparticles: the influence of size, shape, and dielectric environment (2003).

\bibitem{scholl2012quantum}
J.~A. Scholl, A.~L. Koh, J.~A. Dionne, Quantum plasmon resonances of individual metallic nanoparticles, Nature 483~(7390) (2012) 421.

\bibitem{scholl2013observation}
J.~A. Scholl, A.~Garc{\'\i}a-Etxarri, A.~L. Koh, J.~A. Dionne, Observation of quantum tunneling between two plasmonic nanoparticles, Nano Lett. 13~(2) (2013) 564--569.

\bibitem{savage2012revealing}
K.~J. Savage, M.~M. Hawkeye, R.~Esteban, A.~G. Borisov, J.~Aizpurua, J.~J. Baumberg, Revealing the quantum regime in tunnelling plasmonics, Nature 491~(7425) (2012) 574--577.

\bibitem{esteban2012bridging}
R.~Esteban, A.~G. Borisov, P.~Nordlander, J.~Aizpurua, Bridging quantum and classical plasmonics with a quantum-corrected model, Nat. Commun. 3 (2012) 825.

\bibitem{esteban2015classical}
R.~Esteban, A.~Zugarramurdi, P.~Zhang, P.~Nordlander, F.~J. Garc{\'\i}a-Vidal, A.~G. Borisov, J.~Aizpurua, A classical treatment of optical tunneling in plasmonic gaps: extending the quantum corrected model to practical situations, Faraday Discuss. 178 (2015) 151--183.

\bibitem{baumberg2019extreme}
J.~J. Baumberg, J.~Aizpurua, M.~H. Mikkelsen, D.~R. Smith, Extreme nanophotonics from ultrathin metallic gaps, Nat. Mater. 18~(7) (2019) 668--678.

\bibitem{chen2019atomistic}
X.~Chen, P.~Liu, L.~Jensen, Atomistic electrodynamics simulations of plasmonic nanoparticles, J. Phys. D Appl. Phys. 52~(36) (2019) 363002.

\bibitem{teperik2013quantum}
T.~V. Teperik, P.~Nordlander, J.~Aizpurua, A.~G. Borisov, Quantum effects and nonlocality in strongly coupled plasmonic nanowire dimers, Opt. Express 21~(22) (2013) 27306--27325.

\bibitem{zhu2016quantum}
W.~Zhu, R.~Esteban, A.~G. Borisov, J.~J. Baumberg, P.~Nordlander, H.~J. Lezec, J.~Aizpurua, K.~B. Crozier, Quantum mechanical effects in plasmonic structures with subnanometre gaps, Nat. Commun. 7 (2016) 11495.

\bibitem{urbieta2018atomic}
M.~Urbieta, M.~Barbry, Y.~Zhang, P.~Koval, D.~S{\'a}nchez-Portal, N.~Zabala, J.~Aizpurua, Atomic-scale lightning rod effect in plasmonic picocavities: A classical view to a quantum effect, ACS Nano 12~(1) (2018) 585--595.

\bibitem{marinica2012quantum}
D.~Marinica, A.~Kazansky, P.~Nordlander, J.~Aizpurua, A.~G. Borisov, Quantum plasmonics: nonlinear effects in the field enhancement of a plasmonic nanoparticle dimer, Nano Lett. 12~(3) (2012) 1333--1339.

\bibitem{campos2019plasmonic}
A.~Campos, N.~Troc, E.~Cottancin, M.~Pellarin, H.-C. Weissker, J.~Lerm{\'e}, M.~Kociak, M.~Hillenkamp, Plasmonic quantum size effects in silver nanoparticles are dominated by interfaces and local environments, Nat. Phys. 15~(3) (2019) 275--280.

\bibitem{barbry2015atomistic}
M.~Barbry, P.~Koval, F.~Marchesin, R.~Esteban, A.~Borisov, J.~Aizpurua, D.~S{\'a}nchez-Portal, Atomistic near-field nanoplasmonics: reaching atomic-scale resolution in nanooptics, Nano Lett. 15~(5) (2015) 3410--3419.

\bibitem{zhao2008methods}
J.~Zhao, A.~O. Pinchuk, J.~M. McMahon, S.~Li, L.~K. Ausman, A.~L. Atkinson, G.~C. Schatz, Methods for describing the electromagnetic properties of silver and gold nanoparticles, Acc. Chem. Res. 41~(12) (2008) 1710--1720.

\bibitem{jensen2008electronic}
L.~Jensen, C.~M. Aikens, G.~C. Schatz, Electronic structure methods for studying surface-enhanced raman scattering, Chem. Soc. Rev. 37~(5) (2008) 1061--1073.

\bibitem{morton2011theoretical}
S.~M. Morton, D.~W. Silverstein, L.~Jensen, Theoretical studies of plasmonics using electronic structure methods, Chem. Rev. 111~(6) (2011) 3962--3994.

\bibitem{corni2001enhanced}
S.~Corni, J.~Tomasi, Enhanced response properties of a chromophore physisorbed on a metal particle, J. Chem. Phys. 114~(8) (2001) 3739--3751.

\bibitem{ringe2010unraveling}
E.~Ringe, J.~M. McMahon, K.~Sohn, C.~Cobley, Y.~Xia, J.~Huang, G.~C. Schatz, L.~D. Marks, R.~P. Van~Duyne, Unraveling the effects of size, composition, and substrate on the localized surface plasmon resonance frequencies of gold and silver nanocubes: a systematic single-particle approach, J. Phys. Chem. C 114~(29) (2010) 12511--12516.

\bibitem{chen2015atomistic}
X.~Chen, J.~E. Moore, M.~Zekarias, L.~Jensen, Atomistic electrodynamics simulations of bare and ligand-coated nanoparticles in the quantum size regime, Nat. Commun. 6 (2015) 8921.

\bibitem{jensen2009atomistic}
L.~L. Jensen, L.~Jensen, Atomistic electrodynamics model for optical properties of silver nanoclusters, J. Phys. Chem. C 113~(34) (2009) 15182--15190.

\bibitem{chen2018morphology}
X.~Chen, L.~Jensen, Morphology dependent near-field response in atomistic plasmonic nanocavities, Nanoscale (2018).

\bibitem{rossi2015quantized}
T.~P. Rossi, A.~Zugarramurdi, M.~J. Puska, R.~M. Nieminen, Quantized evolution of the plasmonic response in a stretched nanorod, Phys. Rev. Lett. 115~(23) (2015) 236804.

\bibitem{rossi2020hot}
T.~P. Rossi, P.~Erhart, M.~Kuisma, Hot-carrier generation in plasmonic nanoparticles: The importance of atomic structure, ACS Nano 14~(8) (2020) 9963--9971.

\bibitem{kuisma2015localized}
M.~Kuisma, A.~Sakko, T.~P. Rossi, A.~H. Larsen, J.~Enkovaara, L.~Lehtovaara, T.~T. Rantala, Localized surface plasmon resonance in silver nanoparticles: Atomistic first-principles time-dependent density-functional theory calculations, Phys. Rev. B 91~(11) (2015) 115431.

\bibitem{kumawat2025efficient}
R.~L. Kumawat, G.~C. Schatz, Efficient modeling of structural, electronic, and optical properties of silver and gold metal nanoclusters and alloys using optimized scc-dftb parameters, J. Phys. Chem. C 129~(2) (2025) 1348--1361.

\bibitem{d2018density}
S.~D’Agostino, R.~Rinaldi, G.~Cuniberti, F.~Della~Sala, Density functional tight binding for quantum plasmonics, J. Phys. Chem. C 122~(34) (2018) 19756--19766.

\bibitem{chellam2025density}
N.~S. Chellam, S.~Chaudhuri, A.~Ghosal, S.~K. Giri, G.~C. Schatz, Density functional tight-binding enables tractable studies of quantum plasmonics, J. Phys. Chem. C (2025).

\bibitem{liu2022plasmon}
Z.~Liu, M.~B. Oviedo, B.~M. Wong, C.~M. Aikens, Plasmon-induced excitation energy transfer in silver nanoparticle dimers: A real-time tddftb investigation, J. Chem. Phys. 156~(15) (2022).

\bibitem{herring2023recent}
C.~J. Herring, M.~M. Montemore, Recent advances in real-time time-dependent density functional theory simulations of plasmonic nanostructures and plasmonic photocatalysis, ACS Nanoscience Au 3~(4) (2023) 269--279.

\bibitem{chaudhary2024optical}
M.~Chaudhary, H.-C. Weissker, Optical spectra of silver clusters and nanoparticles from 4 to 923 atoms from the tddft+ u method, Nat. Commun. 15~(1) (2024) 9225.

\bibitem{asadi2020td}
N.~Asadi-Aghbolaghi, R.~Ruger, Z.~Jamshidi, L.~Visscher, Td-dft+ tb: An efficient and fast approach for quantum plasmonic excitations, J. Phys. Chem. C 124~(14) (2020) 7946--7955.

\bibitem{hohenester2022nanoscale}
U.~Hohenester, G.~Unger, Nanoscale electromagnetism with the boundary element method, Phys. Rev. B 105~(7) (2022) 075428.

\bibitem{mystilidis2023potential}
C.~Mystilidis, X.~Zheng, A.~Xomalis, G.~A. Vandenbosch, A potential-based boundary element implementation for modeling multiple scattering from local and nonlocal plasmonic nanowires, Adv. Theory Simul. 6~(3) (2023) 2200722.

\bibitem{ciraci2013hydrodynamic}
C.~Cirac{\`\i}, J.~B. Pendry, D.~R. Smith, Hydrodynamic model for plasmonics: a macroscopic approach to a microscopic problem, ChemPhysChem 14~(6) (2013) 1109--1116.

\bibitem{ciraci2016quantum}
C.~Ciraci, F.~Della~Sala, Quantum hydrodynamic theory for plasmonics: Impact of the electron density tail, Phys. Rev. B 93~(20) (2016) 205405.

\bibitem{yan2013green}
W.~Yan, N.~A. Mortensen, M.~Wubs, \href{https://link.aps.org/doi/10.1103/PhysRevB.88.155414}{Green's function surface-integral method for nonlocal response of plasmonic nanowires in arbitrary dielectric environments}, Phys. Rev. B 88 (2013) 155414.
\newblock \href {https://doi.org/10.1103/PhysRevB.88.155414} {\path{doi:10.1103/PhysRevB.88.155414}}.
\newline\urlprefix\url{https://link.aps.org/doi/10.1103/PhysRevB.88.155414}

\bibitem{toscano2015resonance}
G.~Toscano, J.~Straubel, A.~Kwiatkowski, C.~Rockstuhl, F.~Evers, H.~Xu, N.~Asger~Mortensen, M.~Wubs, Resonance shifts and spill-out effects in self-consistent hydrodynamic nanoplasmonics, Nature Commun. 6~(1) (2015) 1--11.

\bibitem{mystilidis2023opensans}
C.~Mystilidis, X.~Zheng, G.~A. Vandenbosch, Opensans: a semi-analytical solver for nonlocal plasmonics, Comput. Phys. Commun. 284 (2023) 108609.

\bibitem{eremin2022discrete}
Y.~Eremin, A.~Doicu, T.~Wriedt, Discrete sources method for modeling of the influence of the non-local effect on the absorption of bimetallic core-shell non-spherical plasmonic nanoparticles, J. Quantit. Spectrosc. Radiat. Transf. 277 (2022) 107994.

\bibitem{zheng2022boundary}
X.~Zheng, C.~Mystilidis, A.~Xomalis, G.~A. Vandenbosch, A boundary integral equation formalism for modeling multiple scattering of light from 3d nanoparticles incorporating nonlocal effects, Adv. Theory Simul. 5~(12) (2022) 2200485.

\bibitem{li2017hybridizable}
L.~Li, S.~Lanteri, N.~A. Mortensen, M.~Wubs, A hybridizable discontinuous galerkin method for solving nonlocal optical response models, Comput. Phys. Commun. 219 (2017) 99--107.

\bibitem{moeferdt2018plasmonic}
M.~Moeferdt, T.~Kiel, T.~Sproll, F.~Intravaia, K.~Busch, Plasmonic modes in nanowire dimers: a study based on the hydrodynamic drude model including nonlocal and nonlinear effects, Phys. Rev. B 97~(7) (2018) 075431.

\bibitem{huber2025computational}
L.~Huber, U.~Hohenester, A computational maxwell solver for nonlocal feibelman parameters in plasmonics, J. Phys. Chem. C 129~(5) (2025) 2590--2598.

\bibitem{draine1994discrete}
B.~T. Draine, P.~J. Flatau, Discrete-dipole approximation for scattering calculations, JOSA A 11~(4) (1994) 1491--1499.

\bibitem{taflove2005computational}
A.~Taflove, S.~C. Hagness, M.~Piket-May, Computational electromagnetics: the finite-difference time-domain method, Elsevier Amsterdam, The Netherlands, 2005.

\bibitem{de2002retarded}
F.~J. Garc{\'\i}a~de Abajo, A.~Howie, Retarded field calculation of electron energy loss in inhomogeneous dielectrics, Phys. Rev. B 65~(11) (2002) 115418.

\bibitem{myroshnychenko2008modeling}
V.~Myroshnychenko, E.~Carb{\'o}-Argibay, I.~Pastoriza-Santos, J.~P{\'e}rez-Juste, L.~M. Liz-Marz{\'a}n, F.~J. Garc{\'\i}a~de Abajo, Modeling the optical response of highly faceted metal nanoparticles with a fully 3d boundary element method, Adv. Mater. 20~(22) (2008) 4288--4293.

\bibitem{mennucci2019multiscale}
B.~Mennucci, S.~Corni, Multiscale modelling of photoinduced processes in composite systems, Nat. Rev. Chem. 3~(5) (2019) 315--330.

\bibitem{bonatti2020frontiers}
L.~Bonatti, G.~Gil, T.~Giovannini, S.~Corni, C.~Cappelli, Plasmonic resonances of metal nanoparticles: Atomistic vs. continuum approaches, Front. Chem. 8 (2020) 340.

\bibitem{marcheselli2020simulating}
J.~Marcheselli, D.~Chateau, F.~Lerouge, P.~Baldeck, C.~Andraud, S.~Parola, S.~Baroni, S.~Corni, M.~Garavelli, I.~Rivalta, Simulating plasmon resonances of gold nanoparticles with bipyramidal shapes by boundary element methods, J. Chem. Theory Comput. 16~(6) (2020) 3807--3815.

\bibitem{coccia2020hybrid}
E.~Coccia, J.~Fregoni, C.~Guido, M.~Marsili, S.~Pipolo, S.~Corni, Hybrid theoretical models for molecular nanoplasmonics, J. Chem. Phys. 153~(20) (2020) 200901.

\bibitem{mortensen2021mesoscopic}
N.~A. Mortensen, Mesoscopic electrodynamics at metal surfaces: —from quantum-corrected hydrodynamics to microscopic surface-response formalism, Nanophotonics 10~(10) (2021) 2563--2616.

\bibitem{stamatopoulou2022finite}
P.~E. Stamatopoulou, C.~Tserkezis, Finite-size and quantum effects in plasmonics: manifestations and theoretical modelling, Opt. Mater. Expr. 12~(5) (2022) 1869--1893.

\bibitem{fitzgerald2016quantum}
J.~M. Fitzgerald, P.~Narang, R.~V. Craster, S.~A. Maier, V.~Giannini, Quantum plasmonics, Proceedings of the IEEE 104~(12) (2016) 2307--2322.

\bibitem{baumberg2022picocavities}
J.~J. Baumberg, Picocavities: a primer, Nano Lett. 22~(14) (2022) 5859--5865.

\bibitem{benz2016single}
F.~Benz, M.~K. Schmidt, A.~Dreismann, R.~Chikkaraddy, Y.~Zhang, A.~Demetriadou, C.~Carnegie, H.~Ohadi, B.~de~Nijs, R.~Esteban, J.~Aizpurua, J.~J. Baumberg, \href{https://www.science.org/doi/abs/10.1126/science.aah5243}{Single-molecule optomechanics in “picocavities”}, Science 354~(6313) (2016) 726--729.
\newblock \href {http://arxiv.org/abs/https://www.science.org/doi/pdf/10.1126/science.aah5243} {\path{arXiv:https://www.science.org/doi/pdf/10.1126/science.aah5243}}, \href {https://doi.org/10.1126/science.aah5243} {\path{doi:10.1126/science.aah5243}}.
\newline\urlprefix\url{https://www.science.org/doi/abs/10.1126/science.aah5243}

\bibitem{giovannini2025electric}
T.~Giovannini, L.~Nicoli, S.~Corni, C.~Cappelli, The electric field morphology of plasmonic picocavities, Nano Letters 25~(27) (2025) 10802--–10808.

\bibitem{giovannini2019classical}
T.~Giovannini, M.~Rosa, S.~Corni, C.~Cappelli, A classical picture of subnanometer junctions: an atomistic drude approach to nanoplasmonics, Nanoscale 11~(13) (2019) 6004--6015.

\bibitem{giovannini2020graphene}
T.~Giovannini, L.~Bonatti, M.~Polini, C.~Cappelli, Graphene plasmonics: Fully atomistic approach for realistic structures, J. Phys. Chem. Lett. 11~(18) (2020) 7595--7602.

\bibitem{giovannini2022we}
T.~Giovannini, L.~Bonatti, P.~Lafiosca, L.~Nicoli, M.~Castagnola, P.~Grobas~Illobre, S.~Corni, C.~Cappelli, Do we really need quantum mechanics to describe plasmonic properties of metal nanostructures?, ACS Photonics 9 (2022) 3025--3034.

\bibitem{lafiosca2021going}
P.~Lafiosca, T.~Giovannini, M.~Benzi, C.~Cappelli, Going beyond the limits of classical atomistic modeling of plasmonic nanostructures, J. Phys. Chem. C 125~(43) (2021) 23848--23863.

\bibitem{tomasi2005quantum}
J.~Tomasi, B.~Mennucci, R.~Cammi, Quantum mechanical continuum solvation models, Chem. Rev. 105~(8) (2005) 2999--3094.

\bibitem{tomasi1999ief}
J.~Tomasi, B.~Mennucci, E.~Cances, The {IEF} version of the {PCM} solvation method: an overview of a new method addressed to study molecular solutes at the {QM} ab initio level, J. Mol. Struct.: THEOCHEM 464~(1) (1999) 211--226.

\bibitem{mennucci1997evaluation}
B.~Mennucci, E.~Canc{\`e}s, J.~Tomasi, {Evaluation of solvent effects in isotropic and anisotropic dielectrics and in ionic solutions with a unified integral equation method: Theoretical bases, computational implementation, and numerical applications}, J. Phys. Chem. B 101~(49) (1997) 10506--10517.

\bibitem{cances1997new}
E.~Cances, B.~Mennucci, J.~Tomasi, A new integral equation formalism for the polarizable continuum model: Theoretical background and applications to isotropic and anisotropic dielectrics, J. Chem. Phys. 107~(8) (1997) 3032--3041.

\bibitem{trugler2012}
U.~Hohenester, A.~Trügler, \href{https://www.sciencedirect.com/science/article/pii/S0010465511003274}{Mnpbem – a matlab toolbox for the simulation of plasmonic nanoparticles}, Comput. Phys. Commun. 183~(2) (2012) 370--381.
\newline\urlprefix\url{https://www.sciencedirect.com/science/article/pii/S0010465511003274}

\bibitem{saad2003iterative}
Y.~Saad, Iterative methods for sparse linear systems, SIAM, 2003.

\bibitem{meurant2020krylov}
G.~Meurant, J.~D. Tebbens, Krylov Methods for Nonsymmetric Linear Systems: From Theory to Computations, Springer, 2020.

\bibitem{saad1986gmres}
Y.~Saad, M.~H. Schultz, {GMRES}: {A} generalized minimal residual algorithm for solving nonsymmetric linear systems, SIAM J. Sci. Stat. Comp. 7~(3) (1986) 856--869.

\bibitem{bonatti2022silico}
L.~Bonatti, L.~Nicoli, T.~Giovannini, C.~Cappelli, In silico design of graphene plasmonic hot-spots, Nanoscale Adv. 4~(10) (2022) 2294--2302.

\bibitem{mayer2007formulation}
A.~Mayer, Formulation in terms of normalized propagators of a charge-dipole model enabling the calculation of the polarization properties of fullerenes and carbon nanotubes, Phys. Rev. B 75~(4) (2007) 045407.

\bibitem{pelton2013introduction}
M.~Pelton, G.~W. Bryant, Introduction to metal-nanoparticle plasmonics, Vol.~5, John Wiley \& Sons, 2013.

\bibitem{neto2009electronic}
A.~H. Castro-Neto, F.~Guinea, N.~M. Peres, K.~S. Novoselov, A.~K. Geim, The electronic properties of graphene, Rev. Mod. Phys. 81~(1) (2009) 109.

\bibitem{fang2013gated}
Z.~Fang, S.~Thongrattanasiri, A.~Schlather, Z.~Liu, L.~Ma, Y.~Wang, P.~M. Ajayan, P.~Nordlander, N.~J. Halas, F.~J. Garc{\'\i}a~de Abajo, Gated tunability and hybridization of localized plasmons in nanostructured graphene, ACS Nano 7~(3) (2013) 2388--2395.

\bibitem{liebsch1993surface}
A.~Liebsch, Surface-plasmon dispersion and size dependence of mie resonance: silver versus simple metals, Phys. Rev. B 48~(15) (1993) 11317.

\bibitem{giovannini2019polarizable}
T.~Giovannini, A.~Puglisi, M.~Ambrosetti, C.~Cappelli, Polarizable qm/mm approach with fluctuating charges and fluctuating dipoles: the qm/fqf$\mu$ model, J. Chem. Theory Comput. 15~(4) (2019) 2233--2245.

\bibitem{nicoli2023fully}
L.~Nicoli, P.~Lafiosca, P.~Grobas~Illobre, L.~Bonatti, T.~Giovannini, C.~Cappelli, Fully atomistic modeling of plasmonic bimetallic nanoparticles: nanoalloys and core-shell systems, Front. Photonics 4 (2023) 1199598.

\bibitem{Trugler}
A.~Tr{\"u}gler, Optical properties of metallic nanoparticles, Springer, 2011.

\bibitem{miertuvs1981electrostatic}
S.~Miertu{\v{s}}, E.~Scrocco, J.~Tomasi, Electrostatic interaction of a solute with a continuum. a direct utilizaion of ab initio molecular potentials for the prevision of solvent effects, Chem. Phys. 55~(1) (1981) 117--129.

\bibitem{cammi1995remarks}
R.~Cammi, J.~Tomasi, Remarks on the use of the apparent surface charges (asc) methods in solvation problems: Iterative versus matrix-inversion procedures and the renormalization of the apparent charges, J. Comput. Chem. 16~(12) (1995) 1449--1458.

\bibitem{purisima1995simple}
E.~O. Purisima, S.~H. Nilar, \href{https://onlinelibrary.wiley.com/doi/abs/10.1002/jcc.540160604}{A simple yet accurate boundary element method for continuum dielectric calculations}, J. Comput. Chem. 16~(6) (1995) 681--689.
\newblock \href {http://arxiv.org/abs/https://onlinelibrary.wiley.com/doi/pdf/10.1002/jcc.540160604} {\path{arXiv:https://onlinelibrary.wiley.com/doi/pdf/10.1002/jcc.540160604}}, \href {https://doi.org/https://doi.org/10.1002/jcc.540160604} {\path{doi:https://doi.org/10.1002/jcc.540160604}}.
\newline\urlprefix\url{https://onlinelibrary.wiley.com/doi/abs/10.1002/jcc.540160604}

\bibitem{plasmonX_github}
plasmonx v1.0, GitHub Repository, \url{https://github.com/plasmonX/plasmonX} (2025).

\bibitem{plasmonx_zenodo}
T.~Giovannini, P.~Grobas~Illobre, P.~Lafiosca, L.~Nicoli, L.~Bonatti, S.~Corni, C.~Cappelli, \href{https://doi.org/10.5281/zenodo.18151754}{plasmonx: an open-source code for nanoplasmonics} (Jan. 2026).
\newblock \href {https://doi.org/10.5281/zenodo.18151754} {\path{doi:10.5281/zenodo.18151754}}.
\newline\urlprefix\url{https://doi.org/10.5281/zenodo.18151754}

\bibitem{cmake}
{Kitware}, {CMake: Cross-Platform Make}, \url{https://cmake.org} (2023).

\bibitem{plasmonX_documentation}
plasmonx manual, GitHub Repository, \url{https://plasmonx.readthedocs.io} (2025).

\bibitem{runtest}
{Kitware}, {CTest: CMake Testing Tool}, \url{https://cmake.org/cmake/help/latest/manual/ctest.1.html} (2023).

\bibitem{wsl}
{Microsoft}, {Windows Subsystem for Linux Documentation}, \url{https://learn.microsoft.com/en-us/windows/wsl/} (2023).

\bibitem{mkl}
{Intel Corporation}, {Intel oneAPI Math Kernel Library (MKL)}, \url{https://www.intel.com/content/www/us/en/developer/tools/oneapi/onemkl.html} (2023).

\bibitem{geom}
P.~Grobas~Illobre, Geom: Nanostructure geometry generator, GitHub Repository, \url{https://github.com/pgrobasillobre/geom} (2025).

\bibitem{rodriguez2025quantum}
F.~E.~Q. Rodriguez, P.~Lafiosca, T.~Giovannini, C.~Cappelli, Quantum dynamics of dissipative polarizable media, Phys. Rev. B 111~(23) (2025) 235144.

\bibitem{lipparini2014scalable}
F.~Lipparini, L.~Lagardere, B.~Stamm, E.~Cances, M.~Schnieders, P.~Ren, Y.~Maday, J.-P. Piquemal, Scalable evaluation of polarization energy and associated forces in polarizable molecular dynamics: I. toward massively parallel direct space computations, J. Chem. Theory Comput. 10~(4) (2014) 1638--1651.

\bibitem{bugeanu2015wavelet}
M.~Bugeanu, R.~Di~Remigio, K.~Mozgawa, S.~S. Reine, H.~Harbrecht, L.~Frediani, Wavelet formulation of the polarizable continuum model. ii. use of piecewise bilinear boundary elements, Phys. Chem. Chem. Phys. 17~(47) (2015) 31566--31581.

\bibitem{hohenester2018making}
U.~Hohenester, Making simulations with the mnpbem toolbox big: Hierarchical matrices and iterative solvers, Comput. Phys. Commun. 222 (2018) 209--228.

\bibitem{hager2010hpc}
G.~Hager, G.~Wellein, Introduction to High Performance Computing for Scientists and Engineers, CRC Press, 2010.

\bibitem{link1999alloy}
S.~Link, Z.~L. Wang, M.~A. El-Sayed, Alloy formation of gold-silver nanoparticles and the dependence of the plasmon absorption on their composition, J. Phys. Chem. B 103~(18) (1999) 3529--3533.

\bibitem{rioux2014analytic}
D.~Rioux, S.~Valli{\`e}res, S.~Besner, P.~Mu{\~n}oz, E.~Mazur, M.~Meunier, An analytic model for the dielectric function of au, ag, and their alloys, Adv. Opt. Mater. 2~(2) (2014) 176--182.

\bibitem{papavassiliou1976surface}
G.~C. Papavassiliou, Surface plasmons in small au-ag alloy particles, J. Phys. F: Met. Phys. 6~(4) (1976) L103.

\bibitem{ristig2015nanostructure}
S.~Ristig, O.~Prymak, K.~Loza, M.~Gocyla, W.~Meyer-Zaika, M.~Heggen, D.~Raabe, M.~Epple, Nanostructure of wet-chemically prepared, polymer-stabilized silver--gold nanoalloys (6 nm) over the entire composition range, J. Mater. Chem. B 3~(23) (2015) 4654--4662.

\bibitem{sorensen2021atomistic}
L.~K. S{\o}rensen, A.~D. Utyushev, V.~I. Zakomirnyi, H.~{\AA}gren, Atomistic description of plasmonic generation in alloys and core shell nanoparticles, Phys. Chem. Chem. Phys. 23~(1) (2021) 173--185.

\bibitem{duan2012nanoplasmonics}
H.~Duan, A.~I. Fern{\'a}ndez-Dom{\'\i}nguez, M.~Bosman, S.~A. Maier, J.~K. Yang, Nanoplasmonics: classical down to the nanometer scale, Nano Lett. 12~(3) (2012) 1683--1689.

\bibitem{vmd}
W.~Humphrey, A.~Dalke, K.~Schulten, {VMD} -- {V}isual {M}olecular {D}ynamics, Journal of Molecular Graphics 14 (1996) 33--38.

\bibitem{lafiosca2023qm}
P.~Lafiosca, L.~Nicoli, L.~Bonatti, T.~Giovannini, S.~Corni, C.~Cappelli, Qm/classical modeling of surface enhanced raman scattering based on atomistic electromagnetic models, J. Chem. Theory Comput. 19~(12) (2023) 3616--3633.

\bibitem{giovannini2025modeling}
T.~Giovannini, S.~G{\'o}mez, C.~Cappelli, Modeling raman spectra in complex environments: From solutions to surface-enhanced raman scattering, J. Phys. Chem. Lett. 16~(12) (2025) 3106--3121.

\bibitem{illobre2025mixed}
P.~Grobas~Illobre, P.~Lafiosca, L.~Bonatti, T.~Giovannini, C.~Cappelli, Mixed atomistic--implicit quantum/classical approach to molecular nanoplasmonics, J. Chem. Phys. 162~(4) (2025).

\bibitem{lafiosca2024real}
P.~Lafiosca, L.~Nicoli, S.~Pipolo, S.~Corni, T.~Giovannini, C.~Cappelli, Real-time formulation of atomistic electromagnetic models for plasmonics, J. Phys. Chem. C 128~(41) (2024) 17513--17525.

\bibitem{nicoli2024atomistic}
L.~Nicoli, S.~Sodomaco, P.~Lafiosca, T.~Giovannini, C.~Cappelli, Atomistic multiscale modeling of colloidal plasmonic nanoparticles, ACS Phys. Chem. Au 4~(6) (2024) 669--678.

\bibitem{darve2000fast}
E.~Darve, The fast multipole method: numerical implementation, J. Comput. Phys. 160~(1) (2000) 195--240.

\end{thebibliography}







\end{document}